\documentclass[aps,pra,notitlepage]{revtex4-1}

\usepackage{graphicx}
\begin{document}

\title{Derivation of the statistics of quantum measurements\\
from the action of unitary dynamics}

\author{Keito Hibino}
\author{Kazuya Fujiwara}
\author{Jun-Yi Wu}
\author{Masataka Iinuma}
\author{Holger F. Hofmann}
\email{hofmann@hiroshima-u.ac.jp}
\affiliation{Graduate School of Advanced Sciences of Matter, Hiroshima University, Kagamiyama 1-3-1, Higashi-Hiroshima 739-8530, Japan}

\begin{abstract}
Quantum statistics is defined by Hilbert space products between the eigenstates associated with state preparation and measurement. The same Hilbert space products also describe the dynamics generated by a Hamiltonian when one of the states is an eigenstate of energy $E$ and the other represents an observable $B$. In this paper, we investigate this relation between the observable time evolution of quantum systems and the coherence of Hilbert space products in detail. It is shown that the times of arrival for a specific value of $B$ observed with states that have finite energy uncertainties can be used to derive the Hilbert space product between eigenstates of energy $E$ and eigenstates of the dynamical variable $B$. Quantum phases and interference effects appear in the form of an action that relates energy to time in the experimentally observable dynamics of localized states. We illustrate the relation between quantum coherence and dynamics by applying our analysis to several examples from quantum optics, demonstrating the possibility of explaining non-classical statistics in terms of the energy-time relations that characterize the corresponding transformation dynamics of quantum systems. 
\end{abstract} 

\maketitle

\section{Introduction}

Quantum physics describes the relation between non-commuting physical properties in terms of quantum coherent superpositions of eigenstates, and these superpositions result in the often times paradoxical statistics observed in experiments on quantum systems. Although the Hilbert space formalism provides a complete description of these non-classical correlations, it is difficult to explain how the notions of state vectors and operators relate to the observable phenomena associated with the dynamics of physical properties in time and space. Nevertheless, it seems to be obvious that classical physics is merely an approximate limit of quantum mechanics, and this expectation is confirmed by the success of so-called semi-classical descriptions, which can be justified more rigorously by defining ``classical'' states as coherent superpositions with minimal uncertainties. The success of such theories in laser physics has resulted in more sophisticated formulations of phase space methods in quantum mechanics \cite{Sch87,Cav91,Las93,Alb02,Del04a,Del04b,Mun06}, and these methods show a fundamental relation between the area of classical phase space and the quantum mechanical phase, where the ratio between the two is given by $\hbar$. However, the definition of a phase space in quantum mechanics is not without problems, since a phase space point is defined by pairs of physical properties that can never be measured jointly. This problem is highlighted by the non-positive values that quasi-probabilities necessarily assign to mathematical definitions of such phase space points, which has in turn fueled a discussion about alternative descriptions of quantum statistics and their time evolution in quantum mechanics \cite{Lun12,Hof12,Sch13a,Hof14a,Hal14,Hof17}. These controversies indicate that phase space fails to provide a satisfactory description of motion in quantum systems, highlighting the need to reconsider the relation between phase space concepts and the description of dynamics in quantum mechanics at a more fundamental level \cite{Sch13b,Bri15,Bau15,Hof14b,Hof16a}. 

In this paper, we will approach the relation between action and quantum coherence from an empirical perspective by considering the experimentally accessible evidence for the Hilbert space structure of quantum mechanics. Specifically, we consider the motion of states prepared with a sufficiently large energy uncertainty to identify an energy dependent time of propagation between an initial condition and a target condition determined by a measurement outcome. Quantum theory determines the arrival time at the target condition through phase coherences between the energy components of the final measurement outcomes. We show that these phase coherences determine the relation between arrival time and energy that can be observed with finite uncertainty states. It is then possible to define the action as the energy integral of the arrival time, and this action appears as a quantum mechanical phase in the energy components of the final state, where $\hbar$ is the fundamental ratio between the action and the quantum mechanical phase. Importantly, the quantum phases result in interference effects whenever there are multiple arrival times for the same final state. This is the origin of energy quantization, since periodicities of $T$ result in destructive interferences except for energy values separated by intervals of $2 \pi \hbar/T$. This description of quantization is consistent with other interference patterns predicted by the arrival time method, indicating the fundamental nature of the relation between the laws of motion and the non-classical effects of quantum coherence. 

In general, the quantum statistics of any set of eigenstates can be derived from the laws of motion observed in the time evolution generated by the corresponding energy operator. Since all physical properties can serve as generators of dynamics and hence as energies of motion, this insight provides a starting point for a more thorough analysis of the relation between unitary transformations and quantum statistics. The analysis presented in the following indicates that quantum interference effects can always be mapped onto a specific transformation generated by the observable defining either the initial or the final state. It is therefore possible to develop a more immediate understanding of the physics of quantum coherence by considering that the inner products of Hilbert space actually express a relation between the two different dynamical processes that result in state preparation and in measurements \cite{Hof14b,Hof16a,Hof16b,Hof16c}. In this relation, the modulations of probability associated with quantum interferences can be understood in terms of propagation time differences between two intersections of the orbits represented by the quantum states. It is thus possible to connect the seemingly abstract features of quantum statistics with the more familiar aspects of the dynamics that can still be observed in the presence of additional uncertainties introduced by a noisy environment.

To illustrate the possibilities of our method, we apply it to optical states, where photon numbers and quadratures can serve as observables and generators. We show that the energy spectrum of quadrature eigenstates can be explained in terms of the small arrival time differences between the quadrature and the zero point of its oscillation and point out similar patterns in two mode multi-photon interferences with nearly equal photon numbers in the input beams. In both cases, we find that small shifts in time can be observed as significant shifts in the quantum coherent phases described by the action, resulting in easily observable changes of measurement probabilities for the associated eigenstates. We can therefore conclude that there is a direct causal connection between the dynamics of quantum systems observed with finite uncertainty states and the subtleties of quantum interference. The present paper thus provides the starting point for a systematic and scientifically meaningful characterization and classification of quantum coherence and the associated non-classical statistics. 

The rest of the paper is organized as follows. In sect. \ref{sec:dynamics} we review the relation between dynamics and quantum coherence in quantum physics and explain how the deterministic relation between energy and time can be obtained from the expectation values of states with a finite energy-time uncertainty. In sect. \ref{sec:action} we expand the energy eigenstates into components associated with different arrival times and identify their quantum phases with the action of the dynamics observed at finite uncertainties. In sect. \ref{sec:qlaws} we explain how quantum interferences between different arrival times at $b$ result in energy quantization for periodic motion and in quantum interferences between different return times observed in the energy distribution of the state $\mid b \rangle$. In sect. \ref{sec:DS} we use these results to explain double slit interference as interferences between the two times at which a scan of momentum generated by the position of detection at the remote screen arrives at the physical property described by the double slit state. In sect. \ref{sec:states} we complete the theoretical reconstruction of energy eigenstates from the dynamics, including an analysis of turning points where two arrival times merge into each other. In sect. \ref{sec:HO} we apply the theory to a harmonic oscillator and show that the results are in good agreement with the textbook solutions of Schr\"odinger's equation. In sect. \ref{sec:timeQI} we identify the effects of quantum interferences in time by analyzing the energy distribution of a position eigenstate of the harmonic oscillator. In sect. \ref{sec:momentum} we consider the relation with phase space methods and the WKB approximation and show that the momentum represents a time-like parameter for the action of forces. In sect. \ref{sec:twomode} we apply the theory to two mode interferences of optical states with a fixed total photon number and show that the arrival time method accurately predicts the suppression and enhancement of individual outcomes in multi-photon interferences. In sect. \ref{sec:stateprep} the implications of the result for general quantum states is discussed with respect to the dynamics of quantum state preparation and measurement. It is pointed out that a definition of phase space points is impossible because the description of distance between the orbits by quantum mechanical phases makes it impossible to identify phase space points as intersections between the orbits. Sect. \ref{sec:stateprep} summarizes the discussion and concludes the paper.

\section{Quantum dynamics and experimental evidence}
\label{sec:dynamics}

Since the days of Newton, the description of motion has been a central concern of physics. However, quantum mechanics has changed this situation by replacing the dynamics of physical properties with a time evolution of probabilities. In fact, the time evolution of individual quantum systems is fundamentally unobservable, since measurements would change the state of the system and thus modify the laws of motion. In the following, we will take up the challenge and formulate a theory of motion that bridges the gap between the quantum mechanical formulation and our intuitive understanding of macroscopic dynamics. 


As a starting point, we consider the time evolution of a physical property $B$ that can be described in terms of the self-adjoint operator $\hat{B}$ with eigenstates of $\mid b \rangle$ and eigenvalues of $B_b$. Quantum mechanics does not allow us to assign any reality to eigenstates or eigenvalues, since the quantum system will usually be described by a superposition of the eigenstates $\mid b \rangle$. It is therefore a non-trivial matter to find a suitable description of the time dependence of $B$. Moreover, the time dependence of $B$ will depend on the initial properties of the system at $t=0$, which can only be given in terms of an initial state $\mid a \rangle$ with non-vanishing uncertainties for properties that define the time dependence of $B$. The closest that we get to a detailed description of motion is an expression for the time evolution of the probability amplitudes associated with the different measurement outcomes $B_b$, where the time dependence of the state can be represented using the energy eigenstates $\mid n \rangle$ with energy eigenvalues of $E_n$. The time dependence is then desribed by a multiplication of each energy component with the appropriate phase factor,
\begin{equation}
\label{eq:motion}
\langle b \mid \hat{U}(t) \mid a \rangle = \sum_n \langle b \mid n \rangle \langle n \mid a \rangle
\exp(-i \frac{E_n t}{\hbar}).
\end{equation}
The merit of this well-established description of quantum dynamics is that it allows us to formally separate the laws of motion from the definition of the initial state $\mid a \rangle$ in terms of the three factors in the sum on the right hand side of the equation. Specifically, the universal laws of motion for the property $B$ are determined completely by the combination of the inner product $\langle b \mid n \rangle$ of the eigenstates of $\hat{B}$ and the eigenstates of energy with the phase factor determined by the product of energy eigenvalues $E_n$ and time $t$. It should therefore be possible to identify the state-independent characteristics of the dynamics with the complex values of the Hilbert space inner products $\langle b \mid n \rangle$, establishing a fundamental relation between the dynamics of $\hat{B}$ and the quantum statistics of eigenvalues $B_b$ in eigenstates $\mid n \rangle$ of energy $E_n$. 

The problem we need to address is the statistical randomness introduced by the choice of $\mid a \rangle$. In Eq.(\ref{eq:motion}), this choice is represented by the energy distribution $|\langle n \mid a \rangle|^2$ and the phase relations between the probability amplitudes $\langle n \mid a \rangle$. As can be seen from the fact that the time dependence is described entirely by phase shifts in the energy eigenstate components of $\mid a \rangle$, the phases of $\langle n \mid a \rangle$ determine a time sensitive property of $\mid a \rangle$. Since this property defines the physical meaning of $t=0$, it is reasonable to define the phases of the energy eigenstates $\mid n \rangle$ in such a way that an equal superposition of all eigenstates with phase zero represent an eigenstate of the property that defines the $t=0$ condition in terms of a quantum state $\mid a_0 \rangle$. However, this definition of the $t=0$ condition means that the condition can only be satisfied if the energy of the system is maximally random. This leads to an important conclusion regarding the description of motion in quantum mechanics. The eigenvalue $B_b$ observed at time $t$ is random because of the uncertainty of energy in the $t=0$ condition, while the laws of motion themselves are completely deterministic and reversible. To identify the state independent laws of motion encoded in the Hilbert space inner products $\langle b \mid n \rangle$, it is necessary to separate the initial randomness from the deterministic laws of motion by considering the uncertainty trade-off between energy and the time at which the $t=0$ condition is satisfied. 

In order to minimize the effects of uncertainties in Eq.(\ref{eq:motion}), we need to select a specific energy expectation value $E$ of the energy distribution $|\langle n \mid a \rangle|^2$ and reduce the energy uncertainty associated with the $t=0$ condition by applying an operation to $\mid a_0 \rangle$ which corresponds to a measurement of energy with outcome $E$ and uncertainty $\delta E$. The resulting state $\mid a(E) \rangle$ then has an energy distribution with an uncertainty of $\delta E$ around the expectation value of $E$, while the $t=0$ condition has been changed by the effects of the measurement. Importantly, the change in the $t=0$ condition can be represented by a unitary time evolution, so that the reduction of $\delta E$ corresponds to the Fourier transform of a coherent superposition of time shifts $t-t^\prime$. This distribution of time shifts can be interpreted as an uncertainty in the definition of $t=0$, and hence as a statistical uncertainty in the propagation time $t$. Mathematically, the effects of a coherent superposition of unitary time evolutions on the energy spectrum of the initial state $\mid a (E) \rangle$ can be expressed by a Fourier transform relation between the amplitudes $\langle n \mid a \rangle$ and the coherent amplitudes $G(t-t^\prime)$ of the time shifts $t-t^\prime$, 
\begin{equation}
\label{eq:spread}
\langle n \mid a(E) \rangle = \int G_E(t-t^\prime) \exp(i \frac{E_n}{\hbar} (t-t^\prime)) dt^\prime.
\end{equation}
The initial state $\mid a(E) \rangle$ in Eq.(\ref{eq:motion}) can thus be expressed by a superposition of temporal displacements of the error free $t=0$ condition, 
\begin{equation}
\label{eq:filter}
\langle b \mid \hat{U}(t) \mid a(E) \rangle = \sum_n \langle b \mid n \rangle 
\int G_E(t-t^\prime) \exp(-i \frac{E_n}{\hbar} t^\prime) dt^\prime.
\end{equation}
Here, the time $t^\prime$ determines the energy dependent phase factor of each energy eigenstate, which means that it effectively describes the time of evolution from the $t=0$ condition represented by equal superpositions of all energy eigenstates. The complex function $G(t-t^\prime)$ therefore gives the probability amplitudes of time shifts $t-t^\prime$ between the measurement time $t$ and the propagation time $t^\prime$ at which the system passed through the $t=0$ condition. Due to the Fourier relation between the reduction in the amplitudes $\langle n \mid a(E) \rangle$ given by Eq.(\ref{eq:spread}) and the randomization of propagation times given by Eq.(\ref{eq:filter}), the energy uncertainty $\delta E$ and the time uncertainty $\delta t$ must satisfy the conventional energy-time uncertainty limit of $\delta E \delta t \geq \hbar/2$. 

The uncertainty of $\hat{B}$ at time $t$ observed with the $t=0$ state $\mid a_0 \rangle$ originates from the dependence of $\hat{B}(t)$ on the energy of the system. It is therefore necessary to choose states with sufficiently low energy uncertainties in order to explore the relation between energy $E$ and propagation time $t$ described by Eq.(\ref{eq:motion}). In the limit of low time uncertainties $\delta t$, a reduction of energy uncertainty results in a reduction of the uncertainty of $\hat{B}(t)$, where the expectation values of $\hat{B}(t)$ will depend on the average energy $E$ of the initial state $\mid a (E) \rangle$. It is possible to investigate this relation experimentally, resulting in a description of the dependence of $\hat{B}$ on energy and time by the expectation values $\langle \hat{B} \rangle(t;E)$. In general, this relation depends on the precise quantum statistics of the specific quantum state $\mid a(E) \rangle$. However, the fact that Eq.(\ref{eq:motion}) encodes the time evolution of the statistics of $\hat{B}$ in the form of Hilbert space inner products $\langle b \mid n \rangle$ suggests that all expectation values $\langle \hat{B} \rangle(t;E)$ can be explained in terms of a single fundamental relation $B(t;E)$, where the effects of the uncertainties in energy and in time on the expectation value of $\hat{B}(t)$ can be determined by error propagation. If the function $B(t;E)$ is sufficiently smooth in its dependence on $E$ and on $t$, the error propagation formula is given by the derivatives of second order,
\begin{equation}
\label{eq:deltacorrect}
\langle \hat{B} \rangle(t;E) \approx B(t;E) + \frac{1}{2} \frac{\partial^2}{\partial t^2}B(t;E)\; \delta t^2 + \frac{1}{2} \frac{\partial^2}{\partial E^2}B(t;E)\; \delta E^2.
\end{equation}
This relation expresses the approximate state dependence of the expectation value $\langle \hat{B} \rangle(t;E)$ for all states with sufficiently low uncertainties in energy and in time as a function of the fundamental state-independent relation $B(t;E)$. This fundamental state-independent relation describes the macroscopic effects of the Hilbert space inner products $\langle b \mid n \rangle$ in Eq.(\ref{eq:motion}) on the experimentally observable dynamics of the quantum system. It is therefore possible to derive the quantum statistics of energy eigenstates $\mid n \rangle$ from the relations between propagation time and energy observed in the presence of small but non-negligible uncertainties.

\section{Action as the link between dynamics and quantum coherence}
\label{sec:action}

If we focus on the time dependence of the amplitude of $\mid b \rangle$ given by Eq.(\ref{eq:motion}), we can see that the absolute value of the amplitude will be maximal when the phases of the different components of energy are all equal. The times at which the probability of ebserving a given outcome $\mid b \rangle$ is particularly high are therefore determined by complex phase factors in the inner products $\langle b \mid n \rangle$, and these phase factors must be responsible for the state-independent time dependence $B(t;E)$ observed for initial states $\mid a (E) \rangle$ with finite uncertainties in energy and in time. Since the value of $\hat{B}$ associated with the eigenstate $\mid b \rangle$ is the eigenvalue $B_b$, we can conjecture that the time at which the probability of a specific outcome $\mid b \rangle$ is maximal should satisfy the condition
\begin{equation}
\label{eq:cond1}
B(t;E)=B_b.
\end{equation}
The solution of this equation gives the arrival times at which the probabilities of $\langle b \mid n \rangle$ in Eq.(\ref{eq:motion}) should be maximal, and this condition can now be related to a phase gradient in energy $E$ consistent with the complex phase factor $\exp(-i E_n t/\hbar)$. However, it is possible that Eq.(\ref{eq:cond1}) has more than one solution because the dynamics of the system return to the same eigenvalue $B_b$ at a number of different times $t_\nu$. In order to account for each of these arrival times $t_\nu$, it is necessary to separate the inner products $\langle b \mid n \rangle$ into a sum of complex amplitudes associated with the different arrival times,
\begin{equation}
\label{eq:cohere}
\langle b \mid n \rangle = \sum_\nu A_\nu(B_b,E_n) \exp\left(i \frac{S_\nu(B_b,E_n)}{\hbar}\right).
\end{equation}   
In this decomposition of the inner products, the action $S_\nu$ that determines the complex phase of each contribution is associated with a specific arrival time $t_\nu$ at which the value of $B$ is equal to the eigenvalue $B_b$. At this arrival time, the phases of the energy components close to the energy eigenvalue $E_n$ line up, so that the relation between the actions $S_\nu$ and the arrival time $t_\nu$ can be given as a relation between two neighboring energy eigenstates,
\begin{equation}
\label{eq:arrival}
S_\nu(B_b,E_n) - E_n t_\nu = S_\nu(B_b,E_{n+1}) - E_{n+1} t_\nu.
\end{equation}
This relation holds under the assumption that the arrival time changes only slowly with energy, so that the difference between $t_\nu(B_b,E_{n+1})$ and $t_\nu(B_b,E_n)$ is much smaller than the average value of $t_\nu$ that is valid for a whole range of energy eigenstates with eigenvalues close to $E$. It is then possible to express the relation between the action and the arrival time by continuous functions of energy with
\begin{equation}
\label{eq:action}
t_\nu(B_b,E) = \frac{\partial}{\partial E} S_\nu(B_b,E).
\end{equation}
The arrival time $t_\nu(B_b,E)$ can be obtained by inverting the state-independent time dependence $B(t;E)$ in Eq.(\ref{eq:cond1}). As shown in Eq.(\ref{eq:deltacorrect}), this time dependence can be determined experimentally by applying error propagation to correct the expectation value dynamics obtained with non-vanishing uncertainties of energy and time. Since $B(t;E)$ describes the relations between the expectation values of quantum states with finite uncertainties it is a continuous function that is not limited to the eigenvalues $B_b$ of $\hat{B}$ or to eigenvalues of energy $E_n$. Consequently, the action $S_\nu(B_b,E)$ is also given by a continuous function of $B$ and $E$ that determines the state-independent relation between energy $E$ and arrival time $t$, where the energy $E$ refers to the energy expectation value of the initial states $\mid a(E) \rangle$. 

The central result of the present analysis is that quantum theory establishes a well-defined relation between the macroscopically observable dependence of motion on energy and time and the coherent decomposition of the Hilbert space inner products $\langle b \mid n \rangle$ given by eq.(\ref{eq:cohere}). Specifically, Eq.(\ref{eq:action}) identifies the quantum phases of components of the inner products $\langle b \mid n \rangle$ with the dynamics of $\hat{B}$ observed under the action of the corresponding Hamiltonian on a minimal uncertainty state. It is therefore possible to reconstruct the $\mid b \rangle$-representation of the energy eigenstates $\mid n \rangle$ from the dynamics of $\langle \hat{B} \rangle$ generated by the corresponding Hamiltonian. Quantum coherence and quantum interference effects can then be predicted based on the laws of motion observed under completely different conditions. What we have described here is essentially a fundamental law of physics that allows us to explain the origin of quantum interference effects in terms of a universal relation between the two limits of control represented by dynamics on the one side and static energy eigenstates on the other.

\section{Prediction of quantum interference from the laws of motion}
\label{sec:qlaws}

Before addressing the problem of quantum coherence, one should recall that the laws of motion are much easier to observe than the detailed statistics of a specific quantum state. As noted above, motion can be described in terms of expectation values, with universal features that are robust against experimental noise. Ultimately, this is the reason why the classical approximation is more accessible and more intuitive than the quantum effects observed only in the limit of unusually high precision. In eq.(\ref{eq:motion}), the quantum coherent statistics of an energy eigenstate emerges when the energy uncertainty of $\mid a \rangle=\mid a(E) \rangle$ approaches zero. This happens when the temporal uncertainty described by $G_E(t-t^\prime)$ in eq.(\ref{eq:spread}) becomes very large. It may be tempting to interpret this randomization of time as a technical problem of control, but quantum theory strongly suggests that it is a much more fundamental effect. Any control of energy with a precision of $\delta E$ requires a temporal spread of $\delta t=\hbar/(2\delta E)$. This spread is not described by statistics, but by coherent superpositions. The difference between classical expectations of ideal control and the actual laws of physics governing interactions between external controls and physical systems emerges in the form of observable effects as soon as the contributions from different times $t_\nu$ start to interfere. Whenever the system returns to the same value of $B=B_b$ at different times $t_\nu$, eq.(\ref{eq:cohere}) predicts interference effects in the probability $|\langle b \mid n \rangle|^2$ that reveal an experimentally accessible aspect of the actions $S_\nu(B\b,E)$ that is incompatible with the classical description of motion. By relating these interference effects to the experimentally observable laws of motion, we can thus achieve a better understanding of the origin of randomness in the process of quantum state preparation. Observable quantum interference effects occur whenever a system returns to the same value of $B$ at different times $t_\nu$ in a quantum coherent randomization of its dynamics. The first and most fundamental consequence of this law of physics is quantization itself. If the motion is periodic with period T, eq.(\ref{eq:cohere}) describes an infinite sum of contributions with time differences of $T$ between the $t_\nu$. The action $S_T$ accumulated in each cycle can be obtained from two times $t_\nu$ with $t_{c2}-t_{c1}=T$. According to eq.(\ref{eq:action}), we find
\begin{equation}
\label{eq:cycle}
T = \frac{\partial}{\partial E} S_T(E).
\end{equation}
Since the infinite sum in eq.(\ref{eq:cohere}) vanishes unless the phase $S_T/\hbar$ is equal to a multiple of $2 \pi$, the energies that emerge for $\delta E \to 0$ must satisfy the relation
\begin{equation}
\label{eq:quant1}
E_n T = S_0 + 2 \pi \hbar n
\end{equation}
and the quantization interval between two energy eigenvalues is given by
\begin{equation}
\label{eq:quant2}
\Delta E_Q = \frac{2 \pi \hbar}{T}.
\end{equation}
Note that this explanation of eigenvalue quantization strongly suggests that the emergence of eigenvalues requires a dynamical randomization generating a time uncertainty that exceeds the period of motion $T$. It may therefore be necessary to think of eigenvalues as contextual values emerging only in the case of specific interaction dynamics, as opposed to the widespread notion of context independent ``elements of reality'' that gives rise to quantum paradoxes \cite{Hof11,Hof15,Hof16c}. 

In addition to the quantization of eigenvalues caused by the periodicity of the dynamics, eq. (\ref{eq:cohere}) also predicts interferences between the different times at which the system arrives at a value of $B=B_b$ during a single period of motion. In this case, there is a non-trivial interference effect caused by the action difference between the two times, 
\begin{equation}
\label{eq:action2}
(t_2(B_b,E)-t_1(B_b,E)) = \frac{\partial}{\partial E} (S_2(B_b,E)-S_1(B_b,E)).
\end{equation} 
This interference effect will appear as a modulation of the probability amplitude $\langle b \mid n \rangle$, where the period $E_{\mathrm{mod.}}$ of the energy dependent modulation is given by the arrival time difference,
\begin{equation}
\label{eq:return}
E_{\mathrm{mod.}} = \frac{2 \pi \hbar}{t_2(B_b,E)-t_1(B_b,E)}. 
\end{equation}
It is therefore possible to read off the arrival time difference at $B_b$ from the interference pattern observed in the energy distribution of the state $\mid b \rangle$. 

Since the probability $P(n|b)=|\langle b \mid n \rangle|^2$ is experimentally observable, it is possible to identify arrival time differences in the quantum statistics of the energy generating the dynamics conditioned by the preparation of the corresponding eigenstate of the dynamical variable. Quantum interferences can thus be understood as signatures of the dynamics generated by the observable detected in the measurement. If the absolute values in eq.(\ref{eq:cohere}) are equal for $t_1$ and $t_2$ ($A_1=A_2=A(B_b,E)$), the probabilities are given by
\begin{equation}
P(n|b) = 2 \Delta E_Q |A(B_b,E)|^2 \left(1+\cos\left(\frac{S_2(B_b,E)-S_1(B_b,E)}{\hbar}\right)\right),
\end{equation} 
where the peak-to-peak energy difference is given by $E_{\mathrm{mod.}}$. The laws of quantum mechanics require that the different arrival times $t_\nu$ of an eigenvalue $B_b$ appear as quantum interferences in the energy statistics conditioned by a preparation of the state $\mid b \rangle$. This relation between statistics and dynamics is universally valid and applies to all physical properties. 

\section{Explanation of double slit interference by the effects of forces on the time-reversed measurement of the particle}
\label{sec:DS}

In order to demonstrate that the action provides a universally valid description of all forms of quantum interference, we can now apply the results of the previous section to the most common textbook example of quantum interference, the particle passing through a double slit. A lot of confusion has arisen from the fact that the origin of coherence between the slits is rarely discussed, mainly because it involves a rather complicated interaction between the particle and the double slit screen. Here, we will sidestep the problem by using the time-reversal symmetry of the problem, so that the double slit state can be identified with the measurement outcome of the dynamical variable $B$. Since we are interested in explaining the Hilbert space inner products $\langle x \mid B=\psi \rangle$ of positions $x$ observed in a plane far away from the double slit, we should consider the dynamics generated by an energy of $E=F x$, where $F$ is an arbitrary force acting on the particle. The state $\mid a \rangle$ that defines $t=0$ is given by an equal superposition of all positions $x$, which means that it is the momentum eigenstate with a momentum of zero. We can therefore solve the dynamics of the system using momentum eigenstates with the momentum eigenvalue given by $p=F t$,
\begin{equation}
\hat{U}(t) \mid a \rangle = \mid Ft \rangle.
\end{equation}  
If the distance between the double slit and the position measurement is $L$ and the longitudinal momentum of the particle is $p_0$, the position at which a particle with transverse momentum $p=Ft$ arrives at the double slit screen is given by $L F t/p_0$, which coincides with the slit positions at $\pm d/2$ for times of $t_\nu=\pm d p_0/(2 L F)$. 

Experimentally, one can time reverse the double slit measurement by sweeping the angle of a Gaussian beam initially focused on the position between the slits ($t=0$) right and left. The beam angle $\theta$ can be related to the dynamics generated by the energy $E=F x$ by $\theta=F t/p_0$. The arrival times at the double slit condition $\psi$ are observed at angles of $\theta=\pm d/2L$, which corresponds to arrival times of $t_\pm=\pm p_0 d/(2 F L)$. The dynamics generated by $x$ therefore results in an energy independent arrival time difference of
\begin{equation}
t_+-t_- = \frac{d}{L} \frac{p_0}{F}.
\end{equation} 
From this time reversed scan of the double slit state $\mid \psi \rangle$, we thus know that the distribution of the energy $E=F x$, and hence the distribution of $x$ given by $|\langle x \mid \psi \rangle|^2$, must exhibit interference fringes with a peak-to-peak difference given by the time difference $t_+-t_-$ as shown in eq.(\ref{eq:return}). In units of position $x$, this peak-to-peak distance is given by 
\begin{equation}
\label{eq:dsreturn}
x_{\mbox{mod.}} = \frac{E}{F} = \frac{2 \pi \hbar}{p_0} \frac{L}{d}. 
\end{equation}
We can therefore predict the periodicity of the double slit pattern from the experimental identification of the slit positions, based on the correct identification of position with energy and deflection angle with time. In particular, the arrival time method indicates that double slit interference must occur if the double slit state is associated with a well-defined observable $\psi$. This is always the case if the state at the slits is maximally defined. The coherence between the slits must therefore be associated with a physical property that is fully determined by the dynamics of quantum state preparation. In the case of the double slit, this property is difficult to identify, which makes it preferable to analyze it in terms of the time-reversed dynamics discussed above.

The main merit of the arrival time explanation of double slit interference is that it does not require any ``dualism'' in the explanation. As noted in the previous section, the interference pattern is a necessary consequence of the dynamics generated by the observable. Rather than being a paradoxical effect, double slit interference is merely a natural aspect of the laws of motion for all particles in the microscopic limit sensitive to small actions. If the state of the particle emerging from the double slits is maximally defined, it is not a random combination of positions but a randomized orbit of a kind of tunneling dynamics between the two slits. In the actual experiment, this orbit originates from the quantum coherence of a well defined transverse momentum (usually close to zero), which results in a simple initial coherence between all positions on the screen. As the particle moves through the double slit screen, its position and momentum distributions get transformed in a highly non-linear manner to produce the double slit orbit $\mid \psi \rangle$. The time-reversed analysis shows that this orbit coherently connects two positions, resulting in the correct prediction of the interference pattern. 

It needs to be recognized that the double slit interference pattern provides microscopic information on the shape of the actual eigenstate-orbit represented by $\mid \psi \rangle$ - information that is not properly conveyed by the description of the state as a superposition of two single slit states. For future progress in quantum research, it will thus be important to correctly identify the dynamics responsible for each specific preparation of a quantum state in order to properly explain the emergence of quantum interference effects observed in the statistics of measurement outcomes.

\section{Derivation of eigenstate statistics}
\label{sec:states}

Since the inner products $\langle b \mid n \rangle$ of quantum states provide a microscopic description of the dynamics of minimal uncertainty states of time and energy, it is possible to derive the quantum coherence of $\langle b \mid n \rangle$ completely from the features of the dynamics given by eq.(\ref{eq:motion}). If we can isolate individual arrival times $t_\nu$, the phases of the contributions in eq.(\ref{eq:cohere}) are given by the actions $S_\nu$. In addition, it is possible to identify the absolute values $A_\nu(B_b,E)$ of the contributions from the relation between dynamics and eigenstates. Here, it is not necessary to consider any effects of quantum coherence, since the results can be derived from the requirement that the total probability of one must be conserved.

From the statistics of the time dependence $B(t;E)$, we can determine the time $Dt$ that the system spends in an interval of length $DB$ around a value of $B_b$,
\begin{equation}
Dt = \left|\frac{\partial}{\partial B} t_\nu(B_b,E) \right| DB.
\end{equation}
This relation describes the flow of probability in $B$ through the interval $DB$ observed at the arrival time $t_\nu$. It therefore quantifies the amount of probability contributed by the energy $E$ to the interval $DB$. The relation with $A_\nu$ can be established through eq.(\ref{eq:filter}), which indicates that $A_\nu$ is a probability amplitude of energy. The Fourier relation between time and energy defined by $G(t-t^\prime)$ requires that the probability density for continuous values of $E$ and $B$ is given by
\begin{equation}
A_\nu (B,E) = \sqrt{\frac{1}{2 \pi \hbar} \left|\frac{\partial}{\partial B} t_\nu(B_b,E)\right|}.
\end{equation}
If $\hat{B}$ is the generator of a periodic transformation, it will be quantized with a quantization interval of $\Delta B_Q$ determined by the periodicity according to eq.(\ref{eq:quant2}). To determine the quantum statistics of eigenstates, it is therefore necessary to consider the dynamics associated with both $E$ and $B$ in a symmetric fashion, where the eigenstates $\mid b \rangle$ and $\mid n \rangle$ represent periodic orbits of the energies $B_b$ and $E_n$, respectively. The probability amplitude given by the inner product of the state vectors represents a quantum statistical ``intersection'' of the two orbits. This symmetry is also reflected by the fact that the contribution of a single intersection $\nu$ can be given by the second derivative of the action, 
\begin{equation}
\label{eq:Vleck}
A_\nu (B,E) = \sqrt{\frac{1}{2 \pi \hbar} \left|\frac{\partial^2}{\partial B \partial E} S_\nu(B,E)\right|}.
\end{equation}
The second derivative of the action is also known as van Vleck determinant, after its derivation from the Jacobean matrix of classical phase space transformations \cite{Vle28}. Here, we obtain the same expression without a phase space model based on a general analysis of dynamics in Hilbert space. The symplectic geometry of phase space is obtained because of the symmetry of $E$ and $B$, which can both serve as generators of their respective dynamics. In the phase space picture, this symmetry is used to eliminate time, replacing the unequal pair $(t;E)$ with the interchangeable phase space coordinates $(B,E)$. Unfortunately, phase space coordinates also create the illusion of static realities by hiding the dynamical relation between energy and time that results in interferences between different times observed in energy eigenstates. It is therefore important to remember that the relation between $E$ and $B$ can only be understood in terms of the dynamics generated by either one of them.

If $B$ is quantized, with discrete eigenvalues of $B_b$, the expression for the amplitude $A_\nu$ must be modified so that the density in $B$ becomes a probability of $B_b$,
\begin{equation}
\label{eq:Bquant}
A_\nu (B_b,E) = \sqrt{\frac{\Delta B_Q}{2 \pi \hbar} \left|\frac{\partial^2}{\partial B \partial E} S_\nu(B,E)\right|},
\end{equation}
where $\Delta B_Q$ is the quantization interval between two consecutive eigenvalues of $\hat{B}$. Note that the resulting factor of $\Delta B_Q/(2 \pi \hbar)$ is the inverse periodicity of the unitary transformation generated by $\hat{B}$, while the second derivative of the action can also be interpreted as the inverse rate of change in $E$ caused by the dynamics generated by $\hat{B}$. $|A_\nu|^2$ therefore describes the probability density of the periodic orbit $B_b$ per energy $dE$. 

We can now find the complete expression of the probability amplitude $\langle b \mid n \rangle$ which describes the interference effects between different intersections of the orbits associated with $b$ and with $n$. If we consider only the set of $t_\nu$ within one period, the probability amplitude can be obtained from the corresponding actions $S_\nu$ by
\begin{equation}
\label{eq:Smetric}
\langle b \mid n \rangle = \sqrt{\frac{\Delta E_Q \Delta B_Q}{2 \pi \hbar}} \sum_\nu 
\sqrt{\left|\frac{\partial^2 S_\nu(B_b,E_n)}{\partial B \partial E}\right|} \exp \left(i \frac{S_\nu(B_b,E_n)}{\hbar}\right).
\end{equation}
For convenience, this result is expressed in its symmetric form with regard to $E$ and $B$. As we saw in the double slit analysis above, this symmetry can be exploited if the dynamics generated by the observable $B$ are known. However, the experimental observation of dynamics assigns very different roles to the generator and to the observable variable. In the light of the present analysis, the practical problem arises from the fact that the randomization of dynamics in quantum state preparation is fundamentally unobservable. For the same reasons, precise quantum measurements must involve an unobservable randomization of the orbit along their target observable. Control of the randomized time parameter can only be achieved by allowing for a corresponding uncertainty of energy. Fortunately, only one such trade-off is needed for a complete reconstruction of the quantum coherence given by the action $S_\nu(B,E)$. Importantly, the arguments $E$ and $B$ in $S$ now refer to expectation values, so that the assumption of a continuous phase space spanned by these two coordinates is actually empirically justified, even if the eigenvalues $E_n$ and $B_b$ are quantized. 

eq.(\ref{eq:Smetric}) explains the quantum coherence of eigenstates $\mid b \rangle$ and $\mid n \rangle$ in terms of discrete and separate arrival times $t_\nu$ at which the orbit described by $\mid n \rangle$ intersects the orbit described by $\mid b \rangle$. However, a more complicated situation arises when the orbits do not intersect but merely touch each other. This happens at a turning point, where two solutions merge and disappear because a minimal energy of $V$ is necessary to achieve stationary action by satisfying $\partial S/\partial E=t$ at the turning point. Quantitatively, the turning point condition can be approximated by the time-energy relation 
\begin{equation}
\label{eq:TP}
t_{\pm} = \pm \sqrt{\gamma (E-V)},
\end{equation}
where the coefficient $\gamma$ can be determined by analyzing the turning point dynamics described by $B(t;E)$. According to eq.(\ref{eq:motion}), the relation between the amplitudes $\langle b \mid n \rangle$ and the time dependent amplitude $\langle b \mid \hat{U}(t) \mid a(E) \rangle$ is described by a Fourier Transform. For sufficiently large energy uncertainties ($\langle n \mid a(E) \rangle$ approximately constant), this means that the same time evolution determines the probability amplitudes 
$\langle b \mid n \rangle$ of all energies close to the turning point. Eq.(\ref{eq:TP}) suggests that this time dependence of the amplitude is given by $\exp(-i t^3/(3 \hbar \gamma))$. For sufficiently large values of $E-V$, it is possible to approximate the Fourier transform by separately solving the Fourier close to the stationary phases at $t_{\pm}$. However, the solutions near $E=V$ require the complete Fourier transform, which is given by the Airy function,
\begin{equation}
\label{eq:Ai}
\langle b \mid n \rangle = \sqrt{\rho_N} \; \mbox{Ai}\left(-\left(\frac{\gamma}{\hbar^2}\right)^{1/3} (E-V)\right).
\end{equation}
The normalization factor $\rho_N$ can be obtained from the limit of high energies using the derivative $\partial t/\partial B$. The result can be given in terms of the gradient of $V$ in $B$,
\begin{equation}
\rho_N = \Delta E_Q \Delta B_Q \left(\frac{\gamma}{\hbar^2}\right)^{2/3} \left|\frac{\partial V}{\partial B}\right|.
\end{equation}
As before, the corresponding probability densities for continuous variables are obtained by omitting the corresponding quantization intervals $\Delta E_Q$ or $\Delta B_Q$. 

The Airy function provides a valid description of the energy dependence of the probability amplitude $\langle b \mid n \rangle$ as long as eq.(\ref{eq:TP}) is a valid approximation of the time-energy relation at the turning point. The arrival time approximation of eq.(\ref{eq:Smetric}) diverges from the result of the Airy function for Airy function arguments greater than $-1$, with relative errors larger than ten percent for arguments greater than $-0.75$. A convenient point to connect the Airy function solution with the arrival time approximation is obtained at $-1.42$, where the results intersect. In terms of the coefficient $\gamma$ in eq.(\ref{eq:TP}), the turning point solution should therefore be used for energies with
\begin{equation}
E < V(B) + 1.42 \left(\frac{\hbar^2}{\gamma}\right)^{1/3}.
\end{equation}
Note that this condition can be understood as a quantitative limit for the onset of tunneling effects in the dynamics of $B$. Specifically, the presence of solutions for energies below $V(B_b)$ at $B=B_b$ will become noticeable when the energy uncertainty of the state is smaller than the energy offset defined by $(\hbar^2/\gamma)^{(1/3)}$. 

The proper connection between the Airy function and the arrival time approximation also shows that the action difference between the solutions $t_+$ and $t_-$ at $t_+=t_-$ is given by $S_+ - S_- = -\hbar \pi/2$. This action difference determines the integration constants for the different solutions of $S_\nu$ and results in the correct quantization condition for the eigenvalues $E_n$. In particular, the ground state condition for dynamics limited by turning points at $B_1$ and $B_2$ can be expressed as
\begin{equation}
\label{eq:GS}
S_+(B_2,E_0)-S_+(B_1,E_0) = -(S_-(B_2,E_0)-S_-(B_1,E_0)) = -\frac{\pi \hbar}{2}.
\end{equation} 
Effectively, the two turning points appear to subtract an action of $\pi \hbar$ from the action enclosed between the turning points, corresponding to a ground state energy of $\pi \hbar/T =\Delta E_Q/2$. 

Eqs.(\ref{eq:Smetric}) and (\ref{eq:Ai}) show how the complete structure of Hilbert space can be derived from the time evolution of the statistics observed with finite energy-time uncertainties. For sufficiently large systems, it is then possible to obtain reliable predictions for the statistics of energy eigenstates from the dynamics, including a prediction of energy quantization and of quantum interference effects in both the statistics of $\hat{B}$ in energy eigenstates, and the statistics of energy in eigenstates of $\hat{B}$. In the second half of the paper, we will take a look at specific examples and show how quantum mechanics connects the highly non-classical statistics of quantum interference with the more familiar laws of motion exhibited by the same physical system under different experimental conditions.

\section{Harmonic oscillations}
\label{sec:HO}

Quantum mechanics originated from the observation that the energies of light appear to be quantized in units of energy given by $\hbar \omega=2 \pi \hbar/T$. In photon detectors, these quanta of energy can be measured directly, essentially providing us with an experimental tool for a highly sensitive measurement of the quantized energies of the harmonic oscillations associated with optical modes. Meanwhile, laser technology also makes it possible to observe the harmonic oscillations of the light field in real time \cite{Rie17}. It is therefore interesting to consider how these seemingly contradictory aspects of harmonic oscillators are related at the microscopic level. Based on the general considerations above, we can now proceed to do so.

Our starting point is the observation of oscillations in the expectation value of a single observable $x$. This observable could be the position of a mechanical object, a quadrature component of the optical field, the dipole moment of a nano particle, the metric of space in a gravitational wave, or any other example of a harmonically oscillating physical property. In all cases, the oscillation can be observed as a time dependent change of the expectation value $\langle \hat{x} \rangle (t)$ given by 
\begin{equation}
\langle \hat{x} \rangle (t;E) = A(E) \cos(\omega t),
\end{equation}
where $t=0$ is arbitrarily chosen to coincide with a maximum of the expectation value. We can now explore the relation between the amplitude $A(E)$ and the energy $E$ of the oscillation. Crude observations suggest a quadratic relation of the form
\begin{equation}
\label{eq:rough}
E \approx \frac{1}{2} k A(E)^2,
\end{equation}
where the constant $k$ is easy to obtain in the limit of large amplitudes $A$. However, more precise measurements of the dynamics need to take into account the uncertainties of energy and of time, as explained in sec, \ref{sec:dynamics} above. Specifically, we need the second derivatives of $x(t;E)$ in time and in energy to compensate the effects of uncertainties in the initial conditions on the expectation value dynamics as shown in eq.(\ref{eq:deltacorrect}). Using the rough estimate provided by eq.(\ref{eq:rough}), we find that
\begin{equation}
\label{eq:HOcorrect}
\langle \hat{x} \rangle (t;E) = x(t;E) \left(1-\frac{\omega^2}{2}\delta t^2 - \frac{1}{8 E^2} \delta E^2 \right).
\end{equation}
We can thus obtain the precise relation $x(t;E)$ between energy and time even though it is impossible to precisely control both aspects of the dynamics simultaneously. 

To simplify the analysis, we can consider the case of equal contributions from time uncertainty and from energy uncertainty, 
\begin{equation}
\delta E = 2 E \omega \delta t.
\end{equation}
Using minimal uncertainty states ($\delta E \delta t = \hbar/2$), this results in an energy uncertainty of
\begin{equation}
\delta E^2 = E \hbar \omega,
\end{equation}
which is equal to the energy fluctuations of coherent states. If such states are used to obtain the expectation value dynamics $\langle \hat{x} \rangle (t;E)$, the corrected energy-time relation is
\begin{equation}
x(t;E) = \left(1 + \frac{\hbar \omega}{4 E} \right) \langle \hat{x} \rangle (t;E).
\end{equation}
It is therefore necessary to increase the contribution of $A(E)$ to the energy by a factor that is itself energy dependent. The result is a correction of the rough relation between expectation value amplitude and energy given by eq.(\ref{eq:rough}). For coherent states of amplitude $A$, this corrected form reads
\begin{equation}
E = \frac{1}{2} k A(E)^2 + \frac{\hbar \omega}{2}.
\end{equation}
Both the energy estimate $E$ and the position estimate $x$ thus include the effects of quantum fluctuations. By compensating these effects, we obtain a deterministic relation between the time dependence of $x$ and the energy $E$, even though the available experimental data is limited by the uncertainties of quantum state preparations. 

Once we have determined the precise relation between energy and time, we can invert it to obtain time as a function of energy,
\begin{equation}
\label{eq:HOtime}
t_\nu(x,E)= \pm \frac{1}{\omega} \arccos \left(\sqrt{\frac{k x^2}{2 E}}\right) + n_{\mbox{cycle}} T.
\end{equation} 
This rather intuitive relation identifies the two arrival times within a cycle and their periodic recurrence. As discussed above, the periodic recurrence results in a quantization of energy, as given by the well known relation $\Delta E_Q = \hbar \omega$. In addition, there are two arrival times within each cycle, one before the maximal value of $x$ is achieved, and the other an equal time after that. 

We can now proceed to derive the precise quantum coherence of the energy eigenstates. For that purpose, we need to fix the quantum phases based on our definition of time zero. Since we identify time zero with a turning point, this is done by assigning an action difference of $-\pi \hbar/2$ to the two solutions at the point where they appear to merge. The result is an action of
\begin{eqnarray}
\label{eq:HOaction}
S_\pm(x,E) &=& \int t_\pm(x,E) dE
\nonumber \\
&=& \pm \left(\frac{E}{\omega} \arccos \left(\sqrt{\frac{k}{2 E}} x\right) - x \sqrt{\frac{k}{2 \omega^2}\left(E-\frac{1}{2}k x^2 \right)} - \frac{\pi}{4}\hbar \right).
\end{eqnarray}
One of the first results we can now obtain is the energy of the ground state. According to eq.(\ref{eq:GS}), the action difference between $x=+\sqrt{2E/k}$ and $x=-\sqrt{2E/k}$ needs to be equal to $\pi\hbar/2$. According to eq.(\ref{eq:HOaction}), this difference is given by $\pi E/\omega$, since the $\arccos$ function at the opposite turning points is $0$ and $\pi$, respectively. We can therefore confirm that the energy of the ground state is $E_0=\hbar \omega/2$.

Between the turning points, the position representation of energy eigenstates is given by eq.(\ref{eq:Smetric}). Using eqs.(\ref{eq:HOtime}) and (\ref{eq:HOaction}), this wave function is given by
\begin{equation}
\label{eq:psi}
\langle x \mid n \rangle = \sqrt{\frac{2}{\pi}
\sqrt{\frac{k}{2E-k x^2}}
} 
\cos \left(\frac{S_+(x,E_n)}{\hbar}\right),
\end{equation}
where $S_-=-S_+$ is explicitly included in the cosine function that represents quantum interferences between the two arrival times $t_\pm$. The approximation used here assumes that the two arrival times are sufficiently distinct to assign separate amplitudes to them. This condition begins to fail at the turning points, where the energy $E$ is close to $(1/2) k x^2$. In this case, a good approximation of the position representation of the energy eigenstates can be obtained using the arrival time estimate (\ref{eq:TP}), with
\begin{equation}
V = \frac{1}{2} k x^2
\end{equation} 
for the turning point potential and
\begin{equation}
\gamma = \frac{2}{\omega^2 k x^2}
\end{equation} 
for the coefficient describing the quantitative time-energy relation. With these relations, we can describe the wave function near the turning points using the Airy function solution of eq.(\ref{eq:Ai}),
\begin{equation}
\label{eq:Ai1}
\langle x \mid n \rangle = \left(\frac{4 k}{\hbar \omega |x|} \right)^{1/6} 
\mbox{Ai}\left(-\left( \frac{2\hbar \omega}{k x^2}\right)^{1/3} \left(\frac{E-\frac{1}{2}kx^2}{\hbar \omega} \right)\right).
\end{equation}
This formula applies to the turning points at $x>0$. For $x<0$, the sign should be flipped for energies of $\hbar \omega (n+1/2)$ with odd $n$, since in those cases, the actions of the two arrival time solutions change by an odd multiple of $\hbar \pi$.   

\begin{figure}[th]
\vspace{-2.5cm}
\begin{picture}(500,520)
\put(0,0){\makebox(480,450){
\scalebox{0.9}[0.9]{
\includegraphics{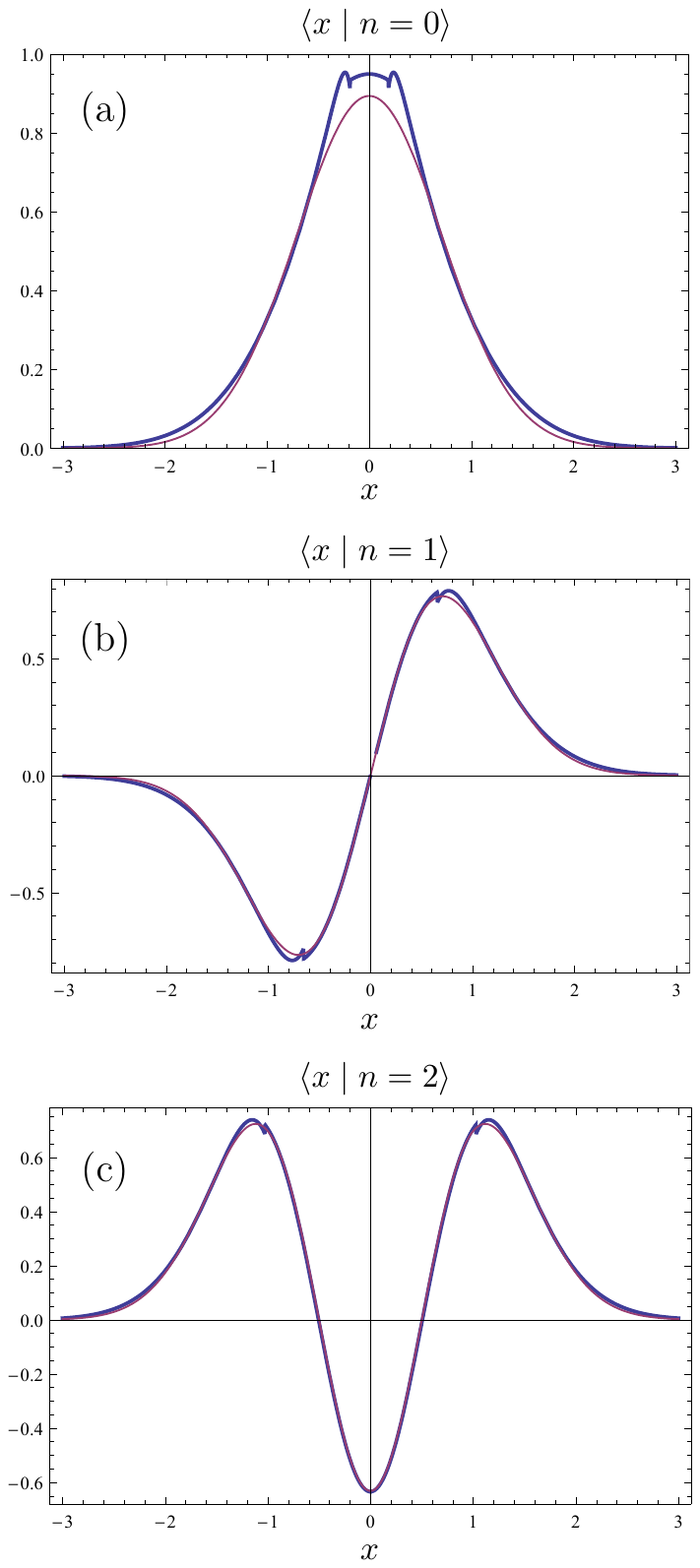}}}}
\end{picture}
\vspace{-0.2cm}
\caption{\label{HOstates}
Comparison of the arrival time solutions for the first three energy eigenstates of the harmonic oscillator (shown by the thick blue lines) with the corresponding solutions of the Schr\"odinger equation (shown by the thin purple lines). (a) shown the amplitudes $\langle x \mid n=0 \rangle $ of the ground state, (b) shows the amplitudes $\langle x \mid n=1 \rangle $ of the first excited state, and (c) shows the amplitudes $\langle x \mid n=2 \rangle $ of the second excited state. Discrepancies are mainly observed in the turning point region, indicating that arrival time provides a good explanation of quantum coherence even at low quantum numbers. 
}
\end{figure}

Eqs.(\ref{eq:psi}) and (\ref{eq:Ai1}) describe the quantum coherence of position and energy in harmonic oscillators based only on the time-energy relation observed in the dynamics of the observable $\hat{x}$. These approximations have been obtained without an operator expression for the Hamiltonian $\hat{H}$ or for the conjugate momentum $\hat{p}$. Nevertheless, the action $S(x,E)$ is in very good agreement with the Gauss-Hermite solutions of the quantum mechanical harmonic oscillator. As readers familiar with phase space methods may have noticed, the result shown in eq.(\ref{eq:psi}) is in fact equal to the result obtained from the Schr\"odinger equation using the WKB approximation for the differential of the wave function in $x$. However, it should be noted that the arrival time method explains the action $S(x,E)$ without introducing a phase space model, and without the use of a conjugate momentum. The arrival time method could thus be used to derive the operator forms of energy and momentum from the experimentally observable dynamics using only the dependence of arrival time on energy. Approximations only enter the derivation of eqs.(\ref{eq:psi}) and (\ref{eq:Ai1}) because the suppression of contributions to $x$ at times different from $t(x,E)$ is only realized asymptotically in eq.(\ref{eq:motion}). However, the suppression of rapidly oscillating phases is so strong that noticeable errors are limited to the turning point regions, where a more detailed analysis would have to focus on the actual relation between energy and time in the region close to the turning point. 

Fig. \ref{HOstates} shows a comparison of the arrival time results with the Gauss-Hermite solutions of the Schr\"odinger equation for the first three states, $n=0$, $n=1$, and $n=2$, where $k=2 \hbar\omega$ has been used to obtain dimensionless values of $x$ consistent with the normalization of quadrature components in quantum optics. As expected, the errors in the ground state are quite noticeable. However, the qualitative features of the ground state Gaussian are approximated rather well, especially when one considers the differences between eq.(\ref{eq:TP}) and eq.(\ref{eq:HOtime}) in this regime. On the other hand, the approximation already works extremely well for $n=1$ and for $n=2$, with the only error occurring at the points where the Airy function solution is used instead of the separate arrival time contributions. These kinks in the graphs show how big the mismatch is between the turning point approximation eq.(\ref{eq:TP}) and the precise description of arrival times by eq.(\ref{eq:HOtime}) at the point where the Airy function argument is $-1.42$. The smallness of these kinks shows that the turning point approximation works reasonably well, even at these low quantum numbers. Aside from these small kinks, the arrival time method correctly explains the shape of the wave functions, indicating that the relation between dynamics and coherence can even explain quantum coherences and quantum interference effects involving only a few quantized energy levels. 

\vspace{1cm}

\section{Temporal quantum interferences in the energy distribution\\ of position eigenstates}
\label{sec:timeQI}

The advantage of explaining quantum coherence in terms of the arrival times $t_\nu$ is that we can learn to relate the patterns of quantum mechanical probabilities to the dynamics generated by the corresponding physical property. As explained in sect. \ref{sec:qlaws}, we can identify eigenvalue differences with periodicities and modulations of probability with arrival time differences. To do so, we should look at the energy distribution of the state $\mid x \rangle$. In quantum optics, this situation could be approximated by a quadrature squeezed state with a non-zero amplitude. For sufficiently high squeezing levels, the photon number distribution will then be given by $P(n)=|\langle x \mid n \rangle|^2$, revealing not just the quantization of photon number by $\Delta E_Q = 2\pi \hbar/T$, but also the modulation period $E_{\mathrm{mod.}}$ associated with the arrival time difference $t_+-t_-$. Since the energy scale is given in terms of photon numbers, it is now particularly easy to relate $t_+-t_-$ to the periodicity in energy, which can be used to identify the position $x$ of the state $\mid x \rangle$. 

\begin{figure}[th]
\vspace{-7cm}
\begin{picture}(500,500)
\put(0,0){\makebox(480,450){
\scalebox{0.85}[0.85]{
\includegraphics{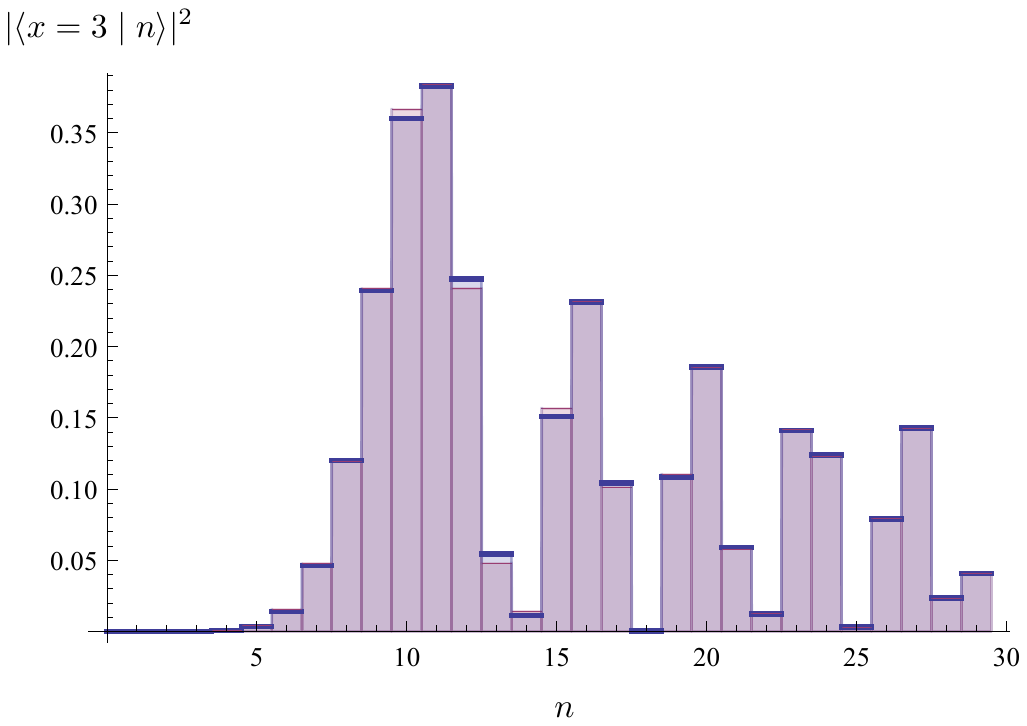}}}}
\end{picture}
\vspace{-4cm}
\caption{\label{x3}
Photon number distribution $|\langle x=3 \mid n \rangle|^2$ of a quadrature eigenstate at $x=3$. Note that probabilities are normalized relative to a finite quadrature interval $dx$ which would correspond to the experimentally achievable squeezing of $x$. The probabilities given by the filled in columns at each photon number show the results of the arrival time method and the thick blue lines indicate exact results obtained from Gauss-Hermite solutions. The arrival time method correctly explains the energy differences between the minima of the distribution.
}
\end{figure}

Fig. \ref{x3} shows the photon number distribution given by $\rho(n)=|\langle x \mid n \rangle|^2$ at $x=3 \sqrt{2(\hbar \omega)/k}$. At this value of $x$, the turning point occurs at $E=9\hbar\omega$, halfway between eight and nine photons. We can now explain the oscillations in photon number probability by considering the difference between the two energy dependent arrival times. According to eq.(\ref{eq:return}), the peak-to-peak distance in energy is equal to $2 \pi \hbar$ divided by the arrival time difference. In the present case, it is easier to use the difference between two minima, since the minima show up more clearly as probabilities close to zero. Although quantization makes it difficult to determine the precise periodicities of quantum interference, we can get a good qualitative idea from the spacing of the three minima at $n=14$ photons, at $n=18$ photons, and at $n=22$ photons. The small non-zero values at $n=14$ and at $n=22$ indicate that the period $E_{\mathrm{mod.}}$ is slightly longer than $4 \hbar \omega$ at $n<18$ and slightly shorter at $n>18$. The experimentally observable probabilities thus indicate that $E_{\mathrm{mod.}}=4 \hbar \omega$ at $n \approx 18$. According to eq.(\ref{eq:return}), this corresponds to an arrival time difference of $T/4$ or arrival times of $t_\pm=\pm T/8$. At these times, the observable $x$ has $1/\sqrt{2}$ times its maximal value. For $n \approx 18$, this results in an estimate of $x=\sqrt{9.25}=3.04$ for the value of $x$ in units of $\sqrt{2(\hbar \omega)/k}$, which is rather close to the actual value of $x=3$ used in the calculation. It is therefore possible to read off the value of $x$ from the oscillations of $P(n|x)=|\langle x \mid n \rangle|^2$. 

Close to the turning point, the arrival time differences are much shorter than half a period and increase rapidly as the energy increases. However, the arrival times always approach $\pm T/4$ when the energy is much greater than the turning point energy of $\hbar \omega x^2$. In the special case of $x=0$, the arrival times of $\pm T/4$ are independent of energy, resulting in a fixed period of $E_{\mathrm{mod.}}=2 \hbar \omega$, which is why the probabilities of odd photon numbers in a quadrature squeezed state at $x=0$ are zero. However, photon quantization makes the statistics extremely sensitive to small differences in arrival times, since they result in phase shifts that accumulate at higher photon numbers. The arrival time method is therefore extremely useful for the explanation of photon number distributions at very small non-zero values of $x$. In this case, the arrival times can be approximated by subtracting the short propagation times between zero and $x$ from the quarter period $T/4$. Since the velocity of motion at low $x$ will be close to the maximal velocity at $x=0$, we can treat this propagation as the propagation of a free particle, where the velocity is proportional to the square root of energy. Using $k$ and $\omega$ to characterize the properties of the harmonic oscillator, the approximate arrival times are given by
\begin{equation}
t_+(x,E) \approx \frac{T}{4} - \sqrt{\frac{k}{2 \omega^2 E}} x. 
\end{equation}
We can now obtain the action without having to solve the complete quantum mechanical problem, simply by integrating the energy dependence of the approximate arrival times. The result reads 
\begin{equation}
S_+(x,E) \approx \frac{\pi}{2 \omega}(E-\frac{\hbar \omega}{2}) - \sqrt{\frac{2 k}{\omega^2} E} \; x,
\end{equation} 
where the non-zero value of the quadrature $x$ reduces the action by a value proportional to the square root of energy. As mentioned above, the $x=0$ part of the action results in phase differences of $0$ or multiples of $2 \pi$ for even photon numbers, and in phase differences of odd multiples of $\pi$ for odd photon numbers. However, the additional phase shift in the quantum interferences at non-zero values of $x$ reduces this phase shift, resulting in an increase in the probability of odd photon numbers and a decrease in the probability of even photon numbers described by
\begin{equation}
|\langle n \mid x \rangle|^2 = \left\{
\begin{array}{ccc}
\frac{2}{\pi} \sqrt{\frac{k}{2 \hbar \omega (n+1/2)}} \left(\cos\left(\sqrt{\frac{2 k}{\hbar \omega} (n+1/2)} \; x\right)\right)^2 &&\mbox{for even $n$}
\\[0.1cm] 
\frac{2}{\pi} \sqrt{\frac{k}{2 \hbar \omega (n+1/2)}} \left(\sin\left(\sqrt{\frac{2 k}{\hbar \omega} (n+1/2)} \; x\right)\right)^2 &&\mbox{for odd $n$}
\end{array}
\right.
\end{equation}
Specifically, we can predict that the first time that the probability of detecting an even photon number drops to zero is at
\begin{equation}
E_1 \approx \left(\frac{\pi}{4}\right)^2 \frac{(\hbar \omega)^2}{\frac{1}{2} k x^2}.
\end{equation}
Oppositely, it is possible to identify $x$ from the energy $E_1$ at which the probabilities of positive photon numbers first approach zero. 

\begin{figure}[th]
\vspace{-2.5cm}
\begin{picture}(500,500)
\put(0,0){\makebox(480,450){
\scalebox{0.9}[0.9]{
\includegraphics{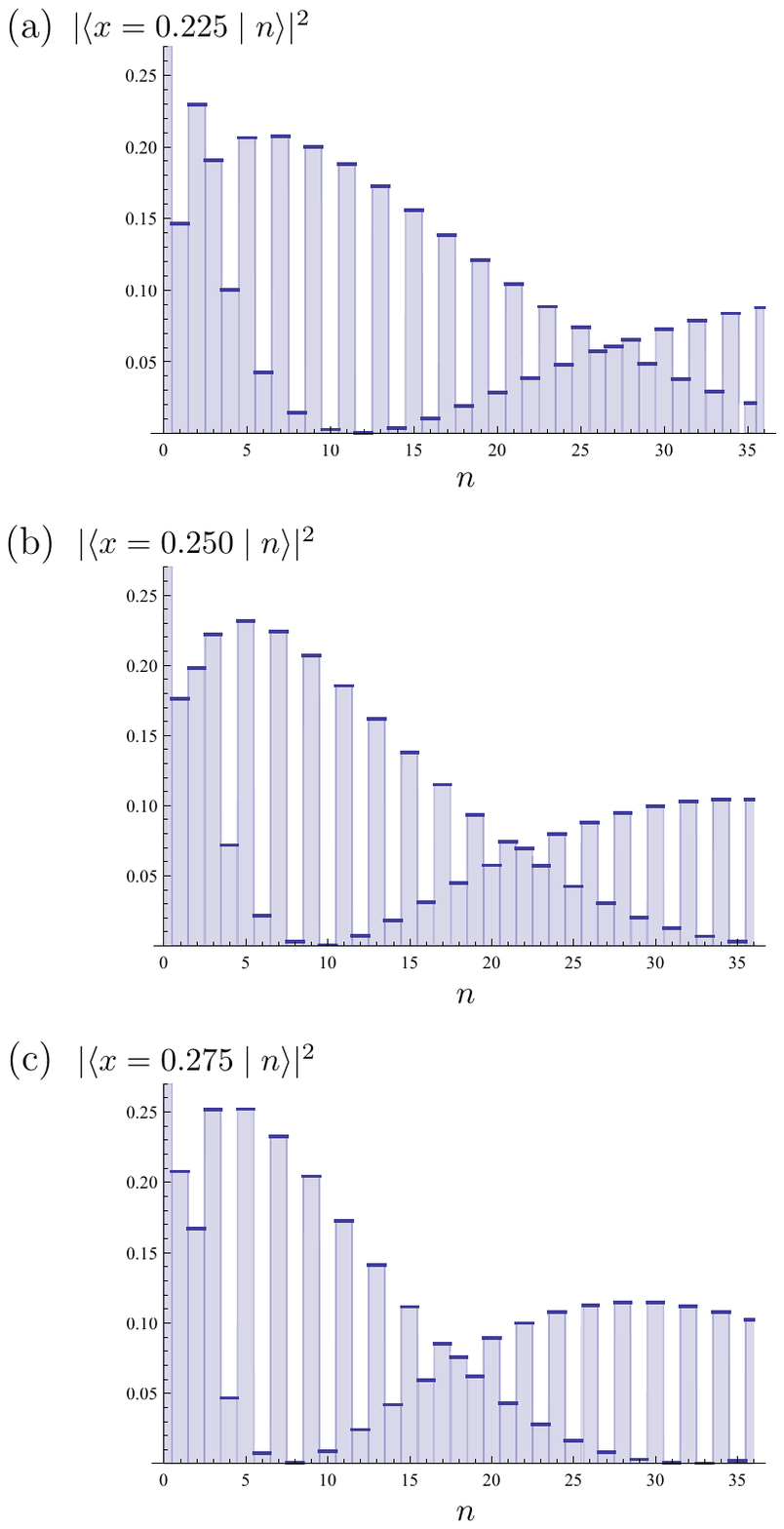}}}}
\end{picture}
\vspace{-0.5cm}
\caption{\label{beats}
Photon number distributions of quadrature eigenstates with (a) $x=0.225$, (b) $x=0.250$, and (c) $x=0.275$. The differences between probabilities of even and odd photon numbers permits a very precise evaluation of the propagation time between $x=0$ and the respective values of $x$, as seen in the gradual shift of interference features towards lower energies.
}
\end{figure}

Fig. \ref{beats} shows the photon number distribution of quadrature eigenstates with three different values of $x$. For the lowest value of $x$, the first minimum of even photon number probability occurs at $n=12$. In units of $\sqrt{2(\hbar \omega)/k}$, this corresponds to a value of $x=0.222$. The actual value for the graph is $x=0.225$, which means that the rough estimate obtained from $E_1$ has an accuracy of 1.3 percent. For the next graph, the minimum occurs between $n=8$ and $n=10$, with a slight bias in favor of the higher photon number. Using a rough estimate of $E_1=\hbar \omega (9.5)$, the estimated value of $x$ is $0.255$. The actual value used for the graph was $0.250$, so the accuracy of the $E_1$ estimate is 2 percent. In the final graph, the minimum has shifted to $n=8$, which gives an estimate of $x=0.270$. The actual value is $x=275$, so the accuracy of the estimate is 1.8 percent. 

If the complete data shown in the graphs was available, a more precise estimate of $x$ would be possible using the features observed at higher photon numbers. However, it is important to note that even the lower photon number results are sensitive to the precise value of $x$, which means that similar results could be obtained using squeezed states with finite average energies. Most importantly, the modulations of probability observed in the photon number statistics are a direct consequence of the distance $x$ between the quadrature eigenstate $\mid x \rangle$ and the center of the oscillation at $x=0$. The phase gradient of the modulation describes the propagation time needed to cover this distance, which means that the distance can be reconstructed from the photon number statistics of the eigenstate using the velocity of the oscillation at $x=0$. 

The example of photon number distributions in quadrature eigenstates illustrates how the arrival time approach to quantum interference can provide better intuitive explanations of quantum interference effects and the associated non-classical statistics of physical properties. It also demonstrates that the method can be applied successfully in the limit of strong quantization, where the discretization of eigenvalues makes it difficult to apply semi-classical concepts of motion and dynamics. It is therefore desirable to develop a better understanding of the relation between the dual roles of physical properties as observables and as generators of dynamics. In textbook quantum mechanics, this role is often assigned to analytical mechanics and to the associated phase space methods. While this may be helpful, it also tends to obscure the different roles played by energy and time in the empirical explanation of the action that we present here. In the next section, we will therefore show how the canonical relation between position and momentum can be explained in terms of experimentally accessible dynamics.

\section{Response to external forces}
\label{sec:momentum}

As mentioned in sect. \ref{sec:HO}, the wave function of the harmonic oscillator states given by eq.(\ref{eq:psi}) corresponds to the results of the WKB approximation for the Schr\"odinger equation of the harmonic oscillator. This correspondence is interesting because we have arrived at the same result without using the Schr\"odinger equation or the operator representation of momentum and energy. The arrival time method can be used to derive the matrix form of the Hamilton operator directly from the dynamics observed at finite uncertainties, without any additional phase space concepts. The only theoretical assumption used is the Hilbert space structure of the unitary time evolution, which seems to represent the essence of quantum theory. 

In the WKB approximation, momentum appears as the derivative of the action assigned to a component of the quantum state. As shown in sect. \ref{sec:dynamics} above, this relation between position and momentum is equivalent to the relation between energy and time. For reasons of consistency, we should therefore consider $p$ as a time-like parameter associated with the unitary transformation
\begin{equation}
\label{eq:up}
\hat{U}(p) = \int \mid x \rangle\langle x \mid \exp(i \frac{x p}{\hbar}) dx.
\end{equation}
It is then possible to analyze the momentum dependence of physical properties by initializing the system in a finite uncertainty state of position and momentum with average momentum of zero defined as equal superposition of all generator eigenstates $\mid x \rangle$. The analogy with time dependence can be explained in terms of a force $F$ applied over a time $t$. The energy is then given by $-Fx$ and the momentum in eq.(\ref{eq:up}) is given by $p=F t$. Experimentally, we can realize the dynamics defined by the momentum $p$ by applying a short ``kick'' to the system, so that the time integral of the force is equal to $p$ and the total time is short enough to neglect all other time evolutions of the system. 

Due to the separation of time scales, it is now possible to investigate the momentum dependence of energy, essentially reversing the roles of generator and observable in the arrival time method. The average energy obtained after applying a ``kick'' of momentum $p$ to a state of finite uncertainty at $x$ can be observed experimentally, e.g. in  quantum optics by photon number measurements on coherent states with the appropriate amplitudes. The statistical correction of the relation between energy $E$ and momentum $p$ at $x$ corresponds to eq.(\ref{eq:deltacorrect}) and is given by
\begin{equation}
\langle \hat{H} \rangle (p;x) = E(p;x) + \frac{1}{2} \frac{\partial^2}{\partial x}E(p; x)\delta x^2 + \frac{1}{2} \frac{\partial^2}{\partial p^2}E(p;x)\delta p^2,
\end{equation}
where $\hat{H}$ is the operator representing the experimentally observable energy $E$. For harmonic oscillators, the uncertainties of position and momentum simply add a constant background energy to the measurement outcomes, with a minimal energy contribution of $\hbar \omega/2$ observed for coherent states. The deterministic relation between energy and time obtained after noise subtraction reads
\begin{equation}
\label{eq:xpenergy}
E (p;x) =\frac{\omega^2}{2 k} p^2 + \frac{k}{2} x^2.
\end{equation}
In textbook quantum mechanics, the Schr\"odinger equation is constructed by replacing the momentum in this relation with the operator $-i \hbar \partial/\partial x$. However, this procedure works only for continuous variables $x$. The time-like parameter $p$ in eq. (\ref{eq:up}) cannot be replaced by an operator, and eq.(\ref{eq:xpenergy}) relates to estimates of $x$, not to eigenvalues. The fundamental relation defining the physical meaning of the time-like parameter $p$ is therefore 
\begin{equation}
\label{eq:momentum}
p(x,E) = \frac{\partial}{\partial x} S(x,E).
\end{equation}
It is somewhat ironic that the WKB approximation uses the same expression to solve the differential equation obtained by replacing the parameter $p$ with an operator $\hat{p}$. 

The main advantage of the operator formalism is that it provides a more detailed description of the turning point regions. To obtain a similar precision with the present approach, we would have to analyze the Fourier relation in eq.(\ref{eq:motion}) in more detail. For reasons of consistency, we can conjecture that this analysis would result in the definition of operators, where the usual representation of the Hamiltonian $\hat{H}$ would be obtained by the matrix representation in $b$. The momentum operator seems to be a rare exception, made possible by the rather unusual fact that momentum shifts of the momentum zero state result in a complete set of orthogonal states. In hindsight, it may be unfortunate that quantum mechanics was historically derived from a description of particle motion, where this replacement of a time-like variable with an operator is accidentally possible. 

In general, it is always possible to exchange the roles of the generator $E$ and the variable $B$ to arrive at an alternative solution of the response to external forces. In fact, it may be better to identify the dynamics with an arbitrary force, where the time-like parameter quantifies the amount of force and the generator observable describes the type of transformation realized by this action. It is then possible to formulate a more consistent description of physical systems in terms of their dynamical structure. 

\section{Photon number statistics of two mode interference}
\label{sec:twomode}

A particularly suitable example of a dynamical structure is provided by the linear optics transformations of light field modes. Light field modes can be represented as a system of harmonic oscillators, where transformations can be described by applying an appropriate frequency difference between two modes. In the case of two modes with photon numbers of $\hat{n}_{a1}$ and $\hat{n}_{a2}$, the unitary operator describing a phase shift of $\phi$ between the two modes is given by
\begin{equation}
\hat{U} = \exp(-i \frac{1}{2}(\hat{n}_{a1}-\hat{n}_{a2}) \phi \rangle.
\end{equation}
Here, the phase $\phi$ is a time-like parameter describing the dynamics of phase shifts and the energy-like generator is given by
\begin{equation}
\hat{J}_1 = \frac{\hbar}{2} \left(\hat{n}_{a1}-\hat{n}_{a2}\right).
\end{equation}
Since linear optics transformations conserve the total photon number, we can focus the discussion on states with a fixed photon number of $n_{a1}+n_{a2}=N$. Phase differences between the modes $a1$ and $a2$ are observed by interfering the two modes at a 50:50 beam splitter, resulting in a photon number difference of $n_{b1}-n_{b2}$ in the output ports of the beam splitter. In the case of optical phase shifts, we thus find that the generator and the dynamical variable are both given by photon number differences. It will be convenient to use the same action scales for both, defining the output intensity difference as
\begin{equation}
\hat{J}_2 = \frac{\hbar}{2} \left(\hat{n}_{b1}-\hat{n}_{b2}\right).
\end{equation}
Note that $J_1$ and $J_2$ are related to energy differences between the fields by a factor of $2 \omega$. The action scale of optical intensities can thus be defined as field energy multiplied by $1/(2 \omega) = T/(4 \pi)$. 

Experimentally observed interference fringes can be described by the amplitude $A$ of the intensity difference $J_2$ in the output of the interferometer. If we define phase zero as the point of maximal intensity difference $J_2$, the experimentally observed phase dependence is
\begin{equation}
\langle \hat{J}_2 \rangle (\phi; J_1) = A(J_1) \cos(\phi).
\end{equation}
Here, the dependence of the amplitude $A$ on $J_1$ results from the fact that an intensity difference between the interfering modes reduces the visibility of the interference fringes. If the total intensity is given by $I$, this relation between $A$ and $J_1$ reads
\begin{equation}
\label{eq:Jrough}
A(J_1) \approx \sqrt{I^2 - J_1^2}.
\end{equation}
We can now determine the correct value of $I$ by investigating the effects of phase uncertainties and generator uncertainties on the observable expectation value dynamics. According to eq.(\ref{eq:deltacorrect}), we have
\begin{equation}
\label{eq:radius1}
\langle \hat{J}_2 \rangle (\phi; J_1) = J_2(\phi; J_1) \left(1 - \frac{1}{2} \delta \phi^2 - \frac{I^2}{2 (I^2-J_1^2)^{2}} \delta J_1^2 \right) 
\end{equation}
If all $N$ photons are prepared in the same optical mode (e.g. by emission from the same coherent source), $\delta J_1^2$ is given by the shot noise of the binomial distribution. In appropriate units, we find $\delta J_1^2 = (\hbar/2) (I^2-J_1^2)/I$. The phase uncertainty of these minimal uncertainty states is given by $\delta \phi^2 = (\hbar/2) I/(I^2-J_1^2)$. This means that the correction terms for $\delta \phi$ and for $\delta J_1$ in eq.(\ref{eq:radius1}) are exactly equal and can be given in the form
\begin{equation}
\label{eq:radius2}
J_2(\phi; J_1) = \langle \hat{J}_2 \rangle (\phi; J_1) \left(1 + \frac{\hbar I}{2 (I^2-J_1^2)} \right). 
\end{equation}
This relation corresponds to a correction of the relation in eq.(\ref{eq:Jrough}), where
\begin{equation}
\label{eq:Jcorrect}
J_1^2 = (I-\hbar/2)^2 - A^2. 
\end{equation}
Since the maximal achievable amplitude is $A_{\mathrm{max.}}=\hbar N/2$, the total intensity $I$ in the deterministic relation $J_2(\phi;J_1)$ is given by
\begin{equation}
I = \frac{\hbar}{2}(N+1).
\end{equation}
Note that the correct identification of the intensity $I$ is essential for a precise determination of quantum coherence in the following theory. It is therefore important that the result can be derived from the statistics of finite uncertainty states, without any speculative theoretical assumptions. It is also interesting to note that the result itself is consistent with the contribution of zero point fluctuations of half a photon from each of the modes. 

Using the correct value of $I$, we can identify the phase shifts $\phi$ needed to arrive at an output intensity difference of $J_2$ at a generator value of $J_1$ as
\begin{equation}
\phi_\nu(J_2,J_1) = \pm \arccos\left(\frac{J_2}{\sqrt{I^2-J_1^2}}\right) + 2 \pi n_{\mbox{cycle}}.
\end{equation}
Since phase shifts are naturally periodic with a period of $2 \pi$, their generators are quantized with a quantization interval of $\Delta J_Q = \hbar$. Within each cycle, we are then left with two arrival phases, one at a negative value before the maximum at $\phi=0$ and one at a positive value after the maximum at $\phi=0$. Using the boundary condition at the turning point, the action that describes the quantum coherence between the two arrival phases is
\begin{eqnarray}
\label{eq:Jact}
\pm S_\pm(J_2,J_1) &=& - J_1 \arccos\left(\frac{J_2}{\sqrt{I^2-J_1^2}}\right) - J_2 \arccos\left(\frac{J_1}{\sqrt{I^2-J_2^2}}\right)
\nonumber \\
&& + I  \arccos\left(\frac{J_1 J_2}{\sqrt{(I^2-J_1^2)(I^2-J_2^2)}}\right) - \frac{\pi}{4} \hbar.
\end{eqnarray}
We can now determine the inner product of eigenstates of $\hat{J}_1$ and eigenstates of $\hat{J}_2$ with quantized eigenvalues of $J_1=\hbar m_1$ and $J_2=\hbar m_2$. Note that the quantum numbers $m_i$ are integers for even photon numbers and half-integers for odd photon numbers, since their values run from $-N/2$ to $+N/2$ in steps of $1$. In the region between the turning points at $J_1^2+J_2^2=I^2$, the inner products of these discrete eigenstates are given by
\begin{equation}
\langle m_2 \mid m_1 \rangle = 2 \sqrt{\frac{\hbar}{2 \pi} \left|\frac{\partial \phi(J_2,J_1)}{\partial J_2} \right|} \; \cos\left( \frac{S_+(J_2,J_1)}{\hbar} \right),
\end{equation}
where the $J_2$ derivative of the phase shift generated by $J_1$ is
\begin{equation}
\frac{\partial \phi(J_2,J_1)}{\partial J_2} = \frac{-1}{\sqrt{I^2-J_1^2-J_2^2}}.
\end{equation}
Note that the result is perfectly symmetric under exchanges of $J_1$ and $J_2$. In fact, the action $S_+(J_s,J_1)$ can also be derived from a phase space description of the spin algebra described by the operators $\hat{J}_1$, $\hat{J}_2$ and a third operator $\hat{J}_3$, such that the two mode $N$ photon states can be identified with points on a sphere \cite{Las93}. In this case, the action is given by the surface area enclosed by circles around $(J_1,0,0)$ and around $(0,J_2,0)$ on a sphere of radius $I$, divided by a normalization factor of $I$. The arrival time method shows how this result can be obtained without any phase space models or semi-classical assumptions, using only the dynamical structure relating phase shifts to their generators. 

\begin{figure}[th]
\vspace{-2.5cm}
\begin{picture}(500,480)
\put(0,0){\makebox(480,420){
\scalebox{0.8}[0.8]{
\includegraphics{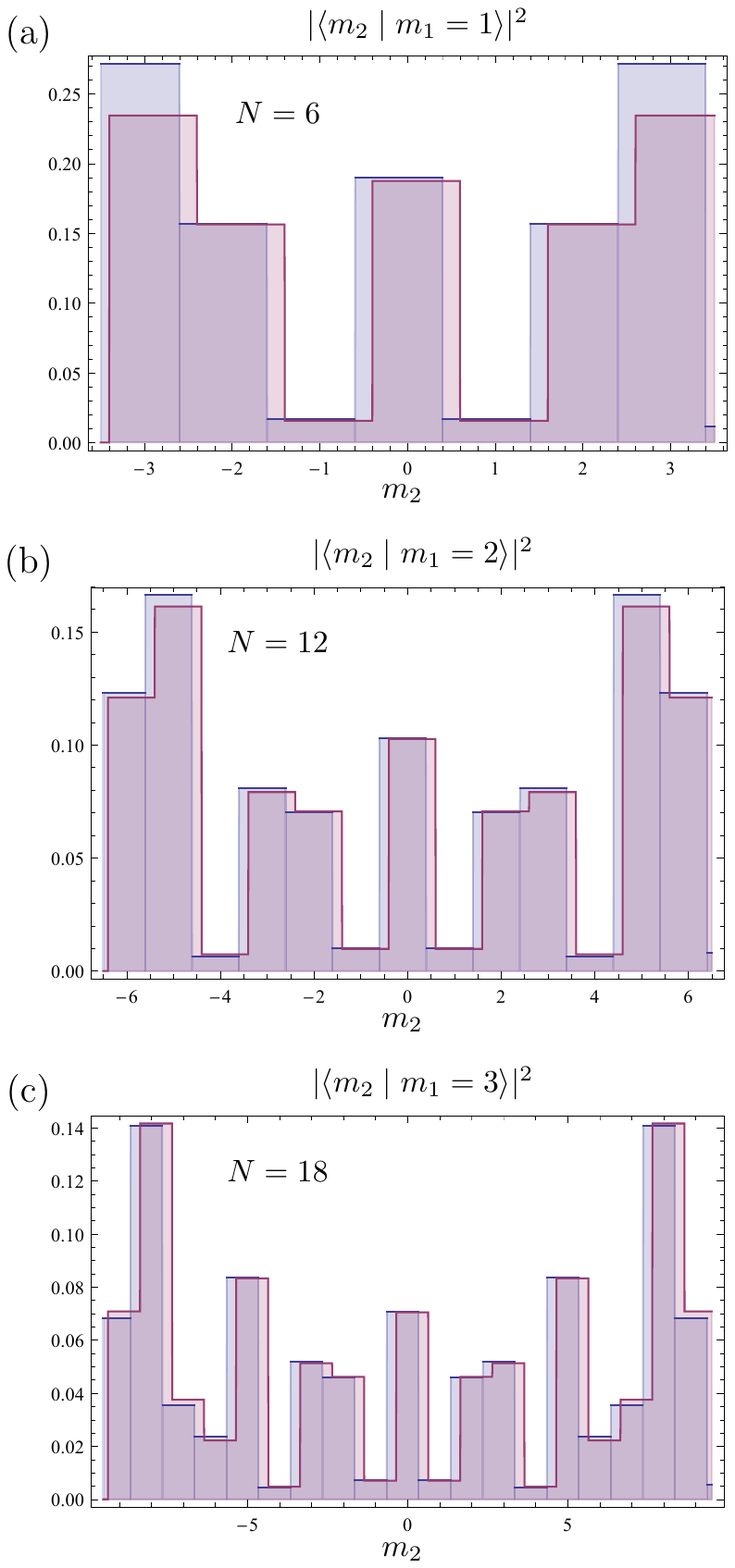}}}}
\end{picture}
\vspace{-0.5cm}
\caption{\label{photons}
Probabilities of observing a photon number difference of $2 m_2$ at the output ports of a beam splitter for input photon number differences of $2 m_1$. (a) shows the probabilities for 6 photons and $m_2=1$, (b) shows the probabilities for 12 photons and $m_2=2$, and (c) shows the probabilities for 18 photons and $m_2=3$. Exact results obtained from the operator algebra (red) are offset slightly to the right and results obtained using the arrival phases (blue) are offset slightly to the left. 
}
\end{figure}

To complete the program, we can now apply the turning point approximation to obtain results for combinations of $J_1$ and $J_2$ near to or beyond the limit of $J_1^2+J_2^2=I^2$. For that purpose, we use the turning point approximation for $\phi$, which is given by 
\begin{equation}
\phi_\pm(J_2,J_1) \approx \pm \sqrt{2 \frac{\sqrt{I^2-J_2^2}}{J_2} \left(\sqrt{I^2-J_2^2} - J_1 \right)}.
\end{equation}
For solutions near $\phi=0$, we then obtain the Airy function approximations
\begin{equation}
\label{eq:AiJ}
\langle m_2 \mid m_1 \rangle = \left(\frac{4 \hbar^2}{J_2\sqrt{I^2-J_2^2}} \right)^{1/6} 
\mbox{Ai}\left(\left(\frac{2 \sqrt{I^2-J_2^2}}{\hbar^2 J_2^2}\right)^{1/3} \left(J_1-\sqrt{I^2-J_2^2} \right)\right).
\end{equation}
Note that this formula is not symmetric under an exchange of $J_1$ and $J_2$. However, the differences are small near the actual turning points. As before, the Airy function approximation should be used when the argument of the Airy function is larger than $-1.42$. We can then obtain a complete set of inner products $\langle m_2 \mid m_1 \rangle$ for the eigenstates of $\hat{J}_1$ and $\hat{J}_2$.

\begin{figure}[th]
\vspace{-2.5cm}
\begin{picture}(500,520)
\put(0,0){\makebox(480,450){
\scalebox{0.85}[0.85]{
\includegraphics{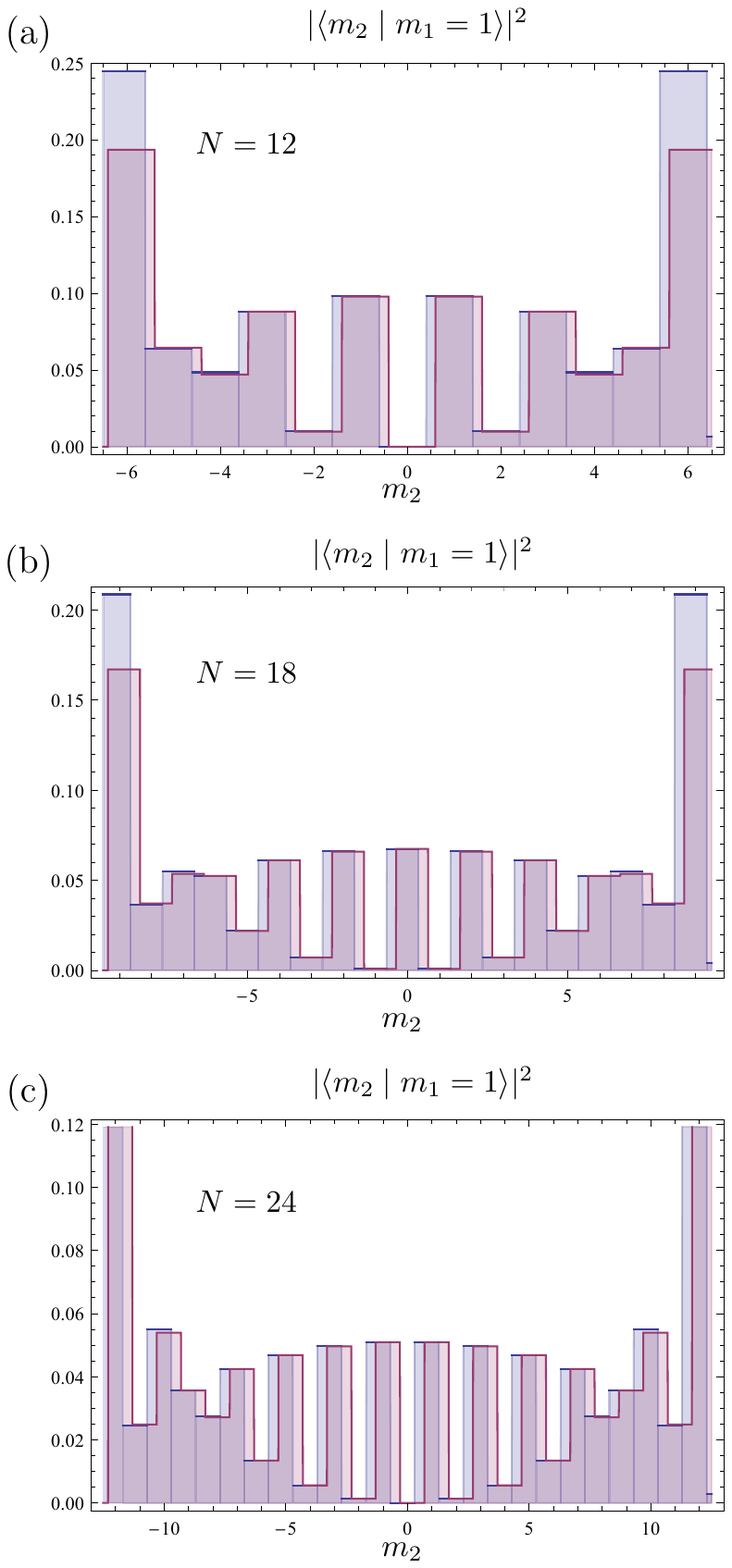}}}}
\end{picture}
\vspace{-0.5cm}
\caption{\label{m1=1}
Probabilities for output photon number differences of $2 m_2$ with input photon number differences of two ($m_1=1$) for (a) $N=12$ photons, (b) $N=18$ photons, and (c) $N=24$ photons. The phase difference between a photon number difference of two and equal photon numbers results in a quadratic $m_2$-dependence of the probability differences between even and odd values of $m_2$. Exact results obtained from the operator algebra (red) are offset slightly to the right and results obtained using the arrival phases (blue) are offset slightly to the left. 
}
\end{figure}

As discussed above, the eigenstates of $\hat{J}_1$ are photon number states of the modes $a1$ and $a2$ with a total photon number of $N$ and a photon number difference of $2 m_1$. The eigenstates of $\hat{J}_2$ are also photon number states, but they describe the photon numbers detected in the modes $b1$ and $b2$, which are obtained by interfering the modes $a1$ and $a2$ at a 50:50 beam splitter. $|\langle m_2 \mid m_1 \rangle|^2$ therefore describes the photon number statistics observed in the output of a beam splitter when the input is given by two photon number states with photon numbers of $N/2+m_1$ and $N/2-m_1$. Fig. \ref{photons} shows the output statistics for 6, 12 and 18 photons with a constant input ratio of $2:1$, compared to the result of an exact analytical calculation using the algebra of creation operators. In all cases, the probabilities of the different photon number distributions are determined by the interference effects between positive and negative phases, and the arrival time method provides reliable predictions for the occurrence of low outcome probabilities. Even at six photons, the suppression of output probabilities for $m_2=\pm 1$ corresponding to 4 photons in one mode and 2 photons in the other is correctly predicted by the action given in eq.(\ref{eq:Jact}).

With regard to suppression effects, it may also be worth noting that the Hong-Ou-Mandel effect can now be explained as a direct consequence of the interference between the arrival phases of $\pi/2$ and of $-\pi/2$ that result in destructive interference for equal output photon numbers. For $m_2=0$ (equal photon numbers in the input), the action is given by 
\begin{eqnarray}
S_+(0,J_1) &=& \frac{\pi}{2}(-J_1 + I - \frac{\hbar}{2}) 
\nonumber \\
&=& \frac{\pi \hbar}{2}(\frac{N}{2} - m_1)
\end{eqnarray}
which results in odd multiples of $\hbar\pi/2$ and hence in destructive interference whenever the difference between $m_1$ and its maximal value is an odd number. Equal photon number outputs are therefore suppressed for odd numbers of pairs $N=2,6,10,\ldots$, while photon differences of two are suppressed for even numbers of pairs $N=4,8,12,\ldots$. The quantitative nature of the phase shift from which the Hong-Ou-Mandel effect arises can be observed by moving to a value of $J_2=\hbar$, or a photon number difference of $2$ in the input. In the dynamical relation between $J_1$ and $J_2$, the arrival phases are shortened by
\begin{equation}
\phi_+(\hbar,J_1) = \frac{\pi}{2}- \frac{\hbar}{\sqrt{I^2-J_1^2}}.
\end{equation}
This small reduction in the arrival phase causes a shift in the action given by
\begin{equation}
S_+(\hbar,J_1) = \frac{\pi \hbar}{2}(\frac{N}{2} - m_1) - \hbar \arccos\left(\frac{J_1}{I}\right).
\end{equation}
At $J_1=0$ (equal photon numbers in the output), the shift is $-\hbar \pi/2$, resulting in output suppression for even numbers of pairs $N=4,8,12,\ldots$, which is the opposite result from the one obtained at $J_2=0$. Near $J_1=I$, the shift of the action returns to zero, while it is $-\hbar \pi$ at $J_1=-I$. The approximate probabilities for $J_2=\hbar$ can be given as
\begin{equation}
\label{eq:m1=1}
|\langle m_1 \mid m_2=1 \rangle|^2 = \left\{
\begin{array}{ccc}
\frac{2}{\pi \sqrt{(N+1)^2/4 - m_1^2}} \left(\frac{m_1^2}{(N+1)^2/4} \right) &&\mbox{for even $N/2-m_1$}
\\[0.1cm] 
\frac{2}{\pi \sqrt{(N+1)^2/4 - m_1^2}} \left(1-\frac{m_1^2}{(N+1)^2/4} \right)&&\mbox{for odd $N/2-m_1$}
\end{array}
\right.
\end{equation}
A simple quadratic function of $m_1$ describes the transition from maximal probability at odd values of $N/2-m_1$ to maximal probability at even values as $m_1$ runs from zero to $N/2$, with approximately equal probabilities for neighboring outcomes near $m_1 \approx (N+1)/(2\sqrt{2})$. This is a significant qualitative change from the probabilities at $m_2=0$, where all $m_1$ with odd values of $N/2-m_1$ were completely suppressed. Three examples of output distributions for input photon number differences of two are shown in fig.\ref{m1=1}. The different probabilities for even and odd values of $m_1$ are clearly visible and correspond well with the prediction of eq.(\ref{eq:m1=1}). The arrival time formalism thus describes the details of multi-photon interferences with a very high level of precision and reliability, even at comparatively low photon numbers.

\section{The role of dynamics in quantum state preparation and measurement}
\label{sec:stateprep}

The arrival time method can be applied to any quantum system with a well-defined dynamical structure given in terms of conjugated pairs of energy and time. It is then possible to derive the quantum coherence of the energy eigenstates $\mid n \rangle$ from the energy dependence of the arrival times $t_\nu$ for the various eigenstates $\mid b \rangle$ of the observable $\hat{B}$. However, the deterministic relation between time and energy cannot be observed directly because of the uncertainties of time and energy. These uncertainties are unavoidable because they originate from the mechanism of quantum state preparation. Any interaction that seeks to reduce the energy uncertainty randomizes the dynamics generated by the energy in a quantum coherent fashion. For fundamental reasons, this randomization cannot be expressed by a random set of unitaries, but must be represented by an integral over the unitary operators that results in a self-adjoint Filter operator for the energy eigenstates. If an energy eigenstate is prepared by this non-classical process of ergodic randomization, interference occurs between different times along the orbit and quantum coherent statics emerge \cite{Hof16a}. For a better understanding of quantum statistics, it is therefore important to recognize the fundamental role of dynamical processes in the control of quantum systems. As the discussion above indicates, phase space pictures fail mostly because they suggest that quantum statistics should be explained in terms of joint realities for non-commuting observables. However, the fundamental relation from which phase space properties emerge are all dynamical, conjugating a parameter describing the amount of change with a single observable property of the system. Quantum interference effects are possible because the time-like parameter describes multiple intersections with the orbits described by other observables. Quantum physics indicates that these intersections do not correspond to joint realities of the physical properties, but need to be evaluated in terms of the complete dynamics of preparation and measurement. Here, it is essential that the state preparation problem is actually the time reverse of the measurement problem. In both instances, the interaction dynamics needed to control an observable $\hat{A}$ with a precision of $\delta A$ coherently randomizes the dynamics generated by $\hat{A}$ to the extent required by the energy-time uncertainty for $\hat{A}$ and its conjugate time $t_A$. An eigenstate $\mid a \rangle$ of $\hat{A}$ can be described as a superposition of eigenstates $\mid b \rangle$ of a different property $\hat{B}$, such that
\begin{equation}
\label{eq:superpose}
\mid a \rangle = \sum_b \langle b \mid a \rangle \mid b \rangle.
\end{equation}
We should now consider the fact that each of the states $\mid b \rangle$ actually represents a complete orbit, and the inner product $\langle b \mid a \rangle$ is not just a joint subset of the two, since its phase is essential for the evaluation of distances between the orbit $\mid b \rangle$ and any other orbits $\mid m \rangle$. These considerations determine the relation between the inner products of Hilbert space, as seen in
\begin{equation}
\label{eq:inner}
\langle m \mid a \rangle = \sum_b \langle b \mid a \rangle \langle m \mid b \rangle.
\end{equation}
We can now derive both $\langle b \mid a \rangle$ and $\langle m \mid b \rangle$ from the dynamics of $\hat{B}$ generated by either $\hat{A}$ or $\hat{M}$, respectively. The inner product $\langle m \mid a \rangle$ is then determined by a sum over the products, representing not only the intersections of $a$ and $b$ and of $b$ and $m$, but also the distance between $a$ and $m$ along $b$ \cite{Hof11,Hof14b}. The present theory presents a more compact map between the observable dynamics and the quantum coherence established when the smallness of the generator uncertainty requires a comparatively large time uncertainty. At this point, the randomness of quantum states cannot be expressed by relations between physical properties any more, and the superposition of eigenstates in eq.(\ref{eq:superpose}) must be seen as a dynamical connection between possibilities. The reason is that inner products do not represent individual points, so that the superposition in eq.(\ref{eq:superpose}) cannot be interpreted as a mere sequence of points $(a,b)$. As a consequence, the inner product $\langle m \mid a \rangle$ that represents a measurement of the orbit $\mid m \rangle$ does not select one of the points $(a,b)$ by identifying it with the condition $(m,b)$. Instead, eq.(\ref{eq:inner}) describes a dynamical sampling process in which the complete orbit $\mid m\rangle$ is compared with the complete orbit $\mid a \rangle$.

Once more, it needs to be emphasized that the dynamics of quantum state preparation and the dynamics of measurement contribute equally to the causality relating the initial conditions to the final outcome. The two conditions are not connected by a static phase space reality. Non-orthogonal quantum states are distinguished by the differences in the dynamics used to generate them. This may be especially important with respect to entangled states, where the correlations between two different systems are defined by non-separable interaction dynamics. The arrival time analysis should be able to shed some light at the emergence of non-classical correlations from the time dependent signatures of such interaction dynamics. It might also be worth noting that there are already some results relating the action phases of unitary transformation to quantum paradoxes, originally motivated by the observation that the weak values of projection operators have phases determined by an action of transformation \cite{Hof11,Hof15}. Based on the detailed analysis presented here, it may be possible to find a more intuitive explanation of the origin of such seemingly paradoxical statistics.

\section{Conclusions}
\label{sec:concl}

In the early days of quantum mechanics, it was practically impossible to observe the dynamics and the quantum statistics of generator eigenstates in the same physical system. This situation has changed, in particular in quantum optics, where the ability to generate a wide variety of non-classical states combines with the large coherent amplitudes of laser light to close the gap between the seemingly classical aspects of continuous quantum dynamics and the fully quantized statistics of photon number distributions. As a result, phase space methods have been particularly successful in quantum optics \cite{Sch87,Cav91,Las93,Alb02,Del04a,Del04b,Mun06,Hof16b}, and the present work could be seen as an attempt to explain the fundamental reasons for this success in more general terms. The main new insight presented here is that the mathematical structure of classical theory can be obtained directly from a quantum theoretical explanation of the empirical evidence, without the use of classical models. It needs to be recognized that the action $S(B,E)$ describes causality relations between initial and final conditions with finite uncertainty. Physical reality only emerges in the course of interactions, where the dynamic control is always limited by energy-time uncertainties. The action $S(B,E)$ quantifies the relation between energy and time and explains the emergence of quantum interference effects as a natural limit of the quantum coherent description of motion. 

Our results raise an interesting question with regard to the controversies concerning the ontological status of physical properties. As we show in the introduction to this paper, the empirical evidence can be explained fully in terms of a quantum theory that strictly limits the control of these properties. Therefore the time evolution $B(t;E)$ cannot be interpreted as a precise trajectory, even though it does describe the deterministic laws of motion valid for the physical system under observation. It seems that there is no good justification for the assumption that $E$ and $B$ must represent elements of reality. Could it be that the desire to attribute reality to physical properties originates from the misconception that causality can only be guaranteed by continuous realities? By sidestepping this philosophical requirement, the present theory manages to smoothly connect classical experience with the fundamental features of quantum theory, successfully identifying the laws of motion as the reason for quantum interference effects and non-classical statistics. It is then possible to gain a better intuitive understanding of the physics behind quantum effects, which will be extremely helpful in the development of larger and more complicated quantum systems and might eventually allow us to unlock the full potential of quantum technologies. 

\section*{Acknowledgements}

This work was supported by JSPS KAKENHI Grant Number 26220712 and by CREST, Japan Science and Technology Agency. KH would like to thank the staff and students of Griffith University for their hospitality and for valuable discussions.


\end{document}